%% file: main.tex
\documentclass[letterpaper,journal]{IEEEtran}
\input{config/header.tex}
\usepackage{fancyhdr}

\fancypagestyle{firstpage}{
  \fancyhf{}
  \fancyhead[C]{\footnotesize
  Submitted to IEEE/ACM TASLP for possible publication. Copyright may be transferred without notice, after which this version may no longer be accessible.}
  
}
\begin{document}
\title{DriftSE: Speech Enhancement with Generative Drifting}
\input{config/authors.tex}%
\maketitle
\thispagestyle{firstpage}
\begin{abstract}
We propose DriftSE, a novel one-step generative framework for speech enhancement formulated as a latent distribution equilibrium problem. During training, the drifting field aligns the generator's pushforward distribution with the clean speech manifold through drifting in a latent domain. During inference, the drifting process is discarded, enabling one-step generation. We establish that its enhancement quality depends fundamentally on the choice of latent representation. Semantic latents preserve phonetic structure but fail to capture physical acoustic cues, whereas acoustic latents reconstruct the physical signal but risk linguistic hallucination. Therefore, we introduce dual-latent drifting, performing parallel drifting in both semantic and acoustic latents to simultaneously preserve phonetic intelligibility and acoustic fidelity. Additionally, we demonstrate that DriftSE enables fully unpaired training by aligning latent distributions rather than exact point-wise targets. Consequently, DriftSE facilitates cross-dataset learning in the absence of paired noisy-clean samples. Moreover, DriftSE exhibits broad architectural flexibility across different generator backbones. Extensive evaluations on additive denoising and convolutive dereverberation demonstrate robust one-step enhancement across both offline and real-time causal settings. Notably, DriftSE achieves state-of-the-art word error rates across all four evaluated datasets while strictly operating at 1 NFE. Code and audio examples are available online\footnote{\url{https://github.com/LiangXu123/DriftSE}}.
\end{abstract}

\begin{IEEEkeywords}
speech enhancement, speech dereverberation, generative models, drifting models
\end{IEEEkeywords}

\input{sections/introduction.tex}

\input{sections/background.tex}

\input{sections/method.tex}

\input{sections/experiments.tex}

\input{sections/results.tex}

\input{sections/conclusion.tex}

\bibliographystyle{IEEEtran}
\bibliography{refs}

\input{config/bios}
\vfill

\end{document}

%% file: config/header.tex
\usepackage{amsmath,amssymb,amsfonts}
\usepackage{xfrac}
\usepackage{array}
\usepackage{bm}
\usepackage{algorithm}
\usepackage{algpseudocode}

\usepackage[table,dvipsnames]{xcolor}
\usepackage{graphicx}
\usepackage{booktabs}
\usepackage{multirow}
\usepackage{tabularx}

\usepackage{float}
\usepackage{cuted}
\usepackage{capt-of}
\usepackage{placeins}
\usepackage{stfloats}

\usepackage{hyperref}
\usepackage[capitalise]{cleveref}
\usepackage{xr}

\usepackage{csquotes}
\usepackage{orcidlink}
\usepackage{pifont}

\usepackage{footnotehyper}
\makesavenoteenv{tabular}
\makesavenoteenv{table}
\usepackage[nolist]{acronym}
\usepackage{silence}
\usepackage{caption}
\usepackage{subcaption}

\usepackage{svg}
\usepackage{varwidth}
\usepackage{tikz}
\usepackage{pgf}
\usepackage{pgfplots}
\pgfplotsset{compat=1.18}
\usepackage{pgfplotstable}
\usetikzlibrary{positioning,backgrounds,fit,calc,patterns,arrows.meta}
\usetikzlibrary{shapes.geometric}
\usetikzlibrary{shapes.arrows}
\usetikzlibrary{arrows.meta}

\usepackage{microtype}

\definecolor{uhhblue}{RGB}{0,156,209}
\definecolor{uhhgreen}{RGB}{66, 178, 60}
\definecolor{uhhred}{RGB}{226,0,26}
\definecolor{uhhblack}{RGB}{0,0,0}
\definecolor{uhhstone}{RGB}{59,81,91}

\def\sg{\mathop{\mathrm{sg}}\nolimits}

\usepackage{xcolor}

\definecolor{driftred}{HTML}{d62728}
\definecolor{driftblue}{HTML}{1f77b4}

\newcolumntype{L}[1]{%
  >{\begin{lrbox}{\nowrapbox}\begin{minipage}{#1}\raggedright}%
  l%
  <{\end{minipage}\end{lrbox}\usebox{\nowrapbox}}%
}

\begin{acronym}
\acro{sgm}[SGM]{score-based generative model}
\acro{snr}[SNR]{signal-to-noise ratio}
\acro{stft}[STFT]{short-time Fourier transform}
\acro{pr}[PR]{phase retrieval}
\acro{istft}[iSTFT]{inverse short-time Fourier transform}
\acro{sde}[SDE]{stochastic differential equation}
\acro{ode}[ODE]{ordinary differential equation}
\acro{pesq}[PESQ]{Perceptual Evaluation of Speech Quality}
\acro{se}[SE]{speech enhancement}
\acro{tf}[T-F]{time-frequency}
\acro{rir}[RIR]{room impulse response}
\acro{snr}[SNR]{signal-to-noise ratio}
\acro{bwe}[BWE]{bandwidth extension}
\acro{lstm}[LSTM]{long short-term memory}
\acro{polqa}[POLQA]{Perceptual Objective Listening Quality Analysis}
\acro{sdr}[SDR]{signal-to-distortion ratio}
\acro{lsd}[LSD]{log-spectral distance}
\acro{sisdr}[SI-SDR]{scale invariant signal-to-distortion ratio}
\acro{estoi}[ESTOI]{Extended Short-Term Objective Intelligibility}
\acro{drr}[DRR]{direct-to-reverberant ratio}
\acro{nfe}[NFE]{number of function evaluations}
\acro{rtf}[RTF]{real-time factor}
\acro{mos}[MOS]{mean opinion scores}
\acro{ema}[EMA]{exponential moving average}
\acro{visqol}[ViSQOL]{Virtual Speech Quality Objective Listener}
\acro{rk}[RK]{Runge-Kutta}
\acro{svd}[SVD]{singular value decomposition}
\acro{dnn}[DNN]{deep neural network}
\acro{mse}[MSE]{mean squared error}
\acro{se}[SE]{speech enhancement}
\acro{bwe}[BWE]{bandwidth extension}
\acro{fm}[FM]{flow matching}
\acro{cfm}[CFM]{conditional flow matching}
\acro{jfm}[JFM]{joint flow matching}
\acro{wer}[WER]{word error rate}
\acro{flop}[FLOP]{floating-point operation}
\acro{api}[API]{application programming interface}
\acro{vbdmd}[VB-DMD]{VoiceBank-DEMAND}
\acro{ewv2}[EWv2]{EARS-WHAM v2}
\acro{lrk}[LRK]{Learned Runge-Kutta}
\acro{db}[DB]{Diffusion Buffer}
\acro{ourmethod}[SFM]{Stream.FM}
\end{acronym}

%% file: config/authors.tex
\author{Liang Xu\,{\orcidlink{0009-0008-8302-437X}},~\IEEEmembership{Student~Member,~IEEE}, Diego Caviedes-Nozal\,{\orcidlink{0000-0001-6756-3375}}, W. Bastiaan Kleijn\,{\orcidlink{0000-0002-1973-3920}},~\IEEEmembership{Fellow,~IEEE}, \\Longfei Felix Yan\,{\orcidlink{0000-0003-4273-198X}},~\IEEEmembership{Member,~IEEE}, and Rasmus Kongsgaard Olsson

\thanks{Liang Xu and W. Bastiaan Kleijn are with Victoria University of Wellington, New Zealand (e-mail: \{liang.xu,bastiaan.kleijn\}@vuw.ac.nz). Diego Caviedes-Nozal and Rasmus Kongsgaard Olsson are with GN Advanced Science, Denmark (e-mail: \{dcnozal,rkolsson\}@gn.com). Longfei Felix Yan is with Lincoln University (e-mail: felix.yan@lincoln.ac.nz).}} 

%% file: sections/introduction.tex
\section{Introduction}
\label{sec:intro}

\IEEEPARstart{S}{peech} enhancement (SE) aims to recover clean and intelligible speech while preserving talker identity across a wide range of acoustic degradations. The design of practical SE systems is largely determined by the latency requirements of the target application, leading to offline \cite{Guo2026Close, richter2023speech} and streaming \cite{welker2026streamfm, lu2025two} paradigms. Offline frameworks leverage the complete audio signal to maximize restoration fidelity. Conversely, streaming models rely on causal processing to enable real-time interaction.

Over the past few decades, speech enhancement has progressed from classical statistical methods, such as Wiener filtering~\cite{meyer1997multi}, to modern neural-network-based techniques~\cite{wang2018supervised}. Discriminative approaches~\cite{Weninger2015LVA, tan18_interspeech, will2016complex, fu2017raw} can effectively suppress additive noise, but their regression-based objectives often lead to perceptually unnatural artifacts and overly smooth spectrogram patterns that lack the fine details of natural speech. To overcome this, generative models~\cite{pascual2017segan, richter2023speech, welker2026streamfm, wang2025flowse} were introduced. Early approaches utilized GANs~\cite{pascual2017segan} to synthesize realistic speech but often suffered from training instabilities and mode collapse. More recently, Score-based Generative Models (SGMs)~\cite{song2021scorebased, richter2023speech} have established a new state-of-the-art by learning the data distribution of clean speech. SGMs~\cite{richter2023speech} define a forward process that gradually corrupts speech with Stochastic Differential Equations and a reverse-time process that reconstructs clean speech along a continuous sampling trajectory.
While this framework excels at both additive denoising and dereverberation, inference is inherently iterative. Numerically integrating this highly curved reverse trajectory requires a high number of function evaluations (NFEs, typically 10--100), causing prohibitive latency for real-time SE.

Efforts to accelerate generation generally fall into either cascaded frameworks or distillation-based methods. Cascaded approaches~\cite{welker2026streamfm, lemercier2023storm} introduce a predictive method to initialize the generation, reducing the second-stage generation process to as few as four inference steps. Alternatively, distillation-based approaches, such as Consistency Models~\cite{song2023consistency, xu2025rosecd, nishigori2025schrodinger}, and trajectory linearization techniques like Flow Matching~\cite{lipman2023flow, rectflow2023, wang2025flowse}, attempt to compress or straighten the generative path itself. Nevertheless, these approaches remain constrained by continuous trajectory modeling, motivating trajectory-free paradigms that natively enable 1-NFE generation.

Recently, Drifting Models~\cite{deng2026generative} emerged as a powerful paradigm that avoids explicit trajectory tracking by formulating generation as a distribution equilibrium problem. Building on this, DriftSE~\cite{xu2026driftse} applied drifting models to speech enhancement through a frame-wise drifting field defined in a single semantic latent space, which steers the generator's pushforward distribution toward the clean distribution during training. By discarding the drifting field at inference, DriftSE enables native 1-NFE enhancement for offline additive denoising.

In this work, we extend DriftSE beyond single-latent offline denoising to address both additive noise and convolutive reverberation. We make three primary contributions. First, we introduce dual-latent drifting, demonstrating that combining semantic and acoustic representations preserves both phonetic intelligibility and acoustic fidelity. Second, we explore the potential of fully unpaired training with frame-wise latent drifting. Through cross-dataset training, we successfully recover acoustic structure in the absence of paired noisy-clean data. Third, we verify architectural flexibility by training with different generator backbones. Extensive evaluations show that DriftSE generalizes across different backbones, achieving robust one-step speech enhancement in both offline and real-time causal configurations. Notably, with all drifting computations confined to training, DriftSE operates strictly at 1 NFE while delivering new state-of-the-art word error rates (WERs) across all four evaluated datasets under both causal and non-causal backbones.

%% file: sections/background.tex
\section{Background}
\label{sec:Background}

Drifting Models~\cite{deng2026generative} cast generative modeling as a dynamic equilibrium process that steers a pushforward distribution toward the target data distribution. This evolution has been formalized as a continuous Wasserstein gradient flow~\cite{turan2026generative, han2026one}, with its underlying mechanics compared against established generative~\cite{song2019generative,yin2024one} paradigms.

\subsection{Pushforward Distribution}
\label{ssec:pushforward_distribution}

Generative modeling can be formulated as learning a mapping $f_\theta$ that transports a source distribution $p_{\epsilon}$ (e.g., standard Gaussian noise) to a target data distribution $p_{\text{data}}$. Given a sample $\epsilon \sim p_\epsilon$, the generator produces an observation $\mathbf{x} = f_\theta(\epsilon)$, and this mapping induces a pushforward distribution defined by
\begin{equation}
    q_\theta := (f_\theta)_\# p_{\epsilon}.
\end{equation}
The objective of generative modeling is to optimize the parameters $\theta$ such that $q_\theta$ converges to $p_{\text{data}}$. For simplicity, we write $q_\theta$ as $q$ and $p_{\text{data}}$ as $p$ in subsequent sections.

In speech enhancement, noisy speech serves as the source distribution, while
the mapping network transports the generated speech distribution toward the clean speech distribution.

\subsection{Wasserstein Gradient Flow}
\label{ssec:wasserstein_flow}

Generative drifting frames the transport of the generated distribution toward the target data distribution as a continuous evolution of probability mass. Let $q_t$ denote the pushforward distribution at continuous time $t \ge 0$. In the 2-Wasserstein space $\mathcal{P}_2(\mathbb{R}^d)$, this evolution follows the steepest descent of the smoothed Kullback-Leibler (KL) energy functional~\cite{turan2026generative}:
\begin{equation}
    F_{\sigma}[q] := \sigma^{2}D_{\mathrm{KL}}(q_{\sigma} \,\|\, p_{\sigma}),
\end{equation}
where $q_\sigma$ and $p_\sigma$ denote the model and target distributions smoothed by a Gaussian kernel with bandwidth $\sigma$ to prevent infinite KL divergence across disjoint supports. The continuous transport is governed by the optimal transport velocity field $\mathbf{v}_\sigma[q_t] = -\nabla_{W_2} F_\sigma[q_t]$, where $\nabla_{W_2}$ is the Wasserstein gradient of $F_\sigma$.

To simulate this continuous evolution over discrete time steps, the JKO scheme~\cite{jordan1998variational, ambrosio2007gradient, santambrogio20151} provides the rigorous variational discretization, yielding an implicit update 
\begin{equation}
    q_{k+1}^{h} \in \arg\min_{q \in \mathcal{P}_2(\mathbb{R}^d)} \left\{ F_{\sigma}[q] + \frac{1}{2h}W_2^2(q,q_k^{h}) \right\},
    \label{eq_jko_update}
\end{equation}
where $q_k^h$ is the discrete distribution at step $k$, and $h$ represents the step size. This implicit optimization evaluates the transport velocity at the unknown future state $q_{k+1}^h$, making it computationally intractable. 

\subsection{Generative Drifting}
\label{ssec_generative_drifting}

Generative drifting bypasses the intractable JKO limitation with a tractable explicit Euler approximation. Freezing the transport velocity at the current state $q_k^h$ yields the direct pushforward update formulated as
\begin{equation}
    \tilde{q}_{k+1}^{h} = \bigl(I + h\,\mathbf{v}_{\sigma}[q_k^{h}]\bigr)_{\#} q_k^{h},
    \label{eq_explicit_euler}
\end{equation}
where $I$ is the identity map and $\tilde{q}_{k+1}^h$ denotes the approximate successor distribution obtained by shifting the mass of $q_k^h$ by a step size $h$ along the frozen velocity $\mathbf{v}_\sigma[q_k^h]$.

Drifting Models~\cite{deng2026generative} realize the update by training the generator to move each generated sample toward its corresponding transport target in a latent space. Let $\mathbf{z}_k=\Phi\!\left(f_{\theta_k}(\epsilon)\right)$ denote the latent representation of the generated observation under a pretrained encoder $\Phi(\cdot)$. The corresponding latent update is
\begin{equation}
    \mathbf{z}_{k+1}
    =
    \mathbf{z}_{k}
    +
    h\,\mathbf{v}_{\sigma}[q_k^h](\mathbf{z}_k),
    \label{eq_latent_update}
\end{equation}
transporting $\mathbf{z}_k$ along the velocity field evaluated at $q_k^h$. To make this computable, drifting models introduce an empirical drifting field $\widehat{\mathbf{v}}(\mathbf{z}_k)$ computed over sample mini-batches (as defined in Section~\ref{ssec_empirical_field}), optimizing the generator parameters $\theta$ such that the output $\Phi(f_\theta(\epsilon))$ matches the drifted target:
\begin{equation}
    \mathcal{L}_{\mathrm{drift}}(\theta)
    =
    \mathbb{E}_{\epsilon\sim p_\epsilon}
    \left[
        \left\|
            \Phi(f_\theta(\epsilon))
            -
            \sg\!\left(
                \mathbf{z}_k
                +
                h\,\widehat{\mathbf{v}}(\mathbf{z}_k)
            \right)
        \right\|_2^2
    \right],
    \label{eq_base_drift_loss}
\end{equation}
where $\sg(\cdot)$ denotes the stop-gradient operator. In practice, the empirical field absorbs the step size, so $h=1$.

\subsection{The Empirical Drifting Field}
\label{ssec_empirical_field}

The drifting objective in ~\eqref{eq_base_drift_loss} requires an empirical estimator $\widehat{\mathbf{v}}$ that defines a meaningful transport direction and vanishes at equilibrium. To reflect its explicit dependence on the target distribution $p$ and the model's pushforward distribution $q$, this field is denoted as $\widehat{\mathbf{v}}(\mathbf{z}) = \mathbf{V}_{p,q}(\mathbf{z})$. Specifically, a valid drifting field must satisfy the equilibrium condition:
\begin{equation}
    p=q
    \quad \Longrightarrow \quad
    \mathbf{V}_{p,q}(\mathbf{z})=\mathbf{0},
    \qquad \forall \mathbf{z}.
    \label{eq:drift_equilibrium}
\end{equation}
This condition is elegantly satisfied by constructing an antisymmetric field (i.e., $\mathbf{V}_{p,q} = -\mathbf{V}_{q,p}$). Inspired by mean shift theory~\cite{meanshift1995}, Drifting Models introduce
\begin{equation}
    \mathbf{V}_{p,q}(\mathbf{z})
    =
    \mathbf{V}_{p}^{+}(\mathbf{z})
    -
    \mathbf{V}_{q}^{-}(\mathbf{z}),
    \label{eq_drift_decomposition_base}
\end{equation}
where $\mathbf{V}_{p}^{+}$ serves as the {attractive force} pulling $\mathbf{z}$ toward high-density data regions of $p$, and $\mathbf{V}_{q}^{-}$ serves as the {repulsive force} pushing it away from model-clustered regions of $q$.

Evaluating forces $\mathbf{V}_{p}^{+}$ and $\mathbf{V}_{q}^{-}$ in raw waveform space is problematic as Euclidean distance can be dominated by signal energy. It may therefore poorly reflect linguistic or acoustic similarity. To enable meaningful structural transport, the drifting field is instead evaluated within a latent space. Let $\mathbf{z} \in \mathbb{R}^d$ denote an arbitrary query point in the latent space. Furthermore, let $\mathbf{z}^+ = \Phi(\mathbf{y}^+)$ denote the latent representation of a target data sample $\mathbf{y}^+ \sim p$, and $\mathbf{z}^- = \Phi(\mathbf{y}^-)$ denote that of a generated model sample $\mathbf{y}^- \sim q$. Both drifting forces act as kernel-weighted mean shift operators~\cite{meanshift1995}, drifting $\mathbf{z}$ toward the local center of mass in latent space:
\begin{align}
    \mathbf{V}_{p}^{+}(\mathbf{z})
    &=
    \frac{1}{Z_p(\mathbf{z})}
    \mathbb{E}_{\mathbf{y}^{+}\sim p}
    \left[
        k_\tau(\mathbf{z},\mathbf{z}^{+})
        (\mathbf{z}^{+}-\mathbf{z})
    \right], \\
    \mathbf{V}_{q}^{-}(\mathbf{z})
    &=
    \frac{1}{Z_q(\mathbf{z})}
    \mathbb{E}_{\mathbf{y}^{-}\sim q}
    \left[
        k_\tau(\mathbf{z},\mathbf{z}^{-})
        (\mathbf{z}^{-}-\mathbf{z})
    \right],
\end{align}
with local normalizers preventing the drift field from vanishing in low-density regions:
\begin{align}
    Z_p(\mathbf{z})
    &=
    \mathbb{E}_{\mathbf{y}^{+}\sim p}
    \left[
        k_\tau(\mathbf{z},\mathbf{z}^{+})
    \right], \\
    Z_q(\mathbf{z})
    &=
    \mathbb{E}_{\mathbf{y}^{-}\sim q}
    \left[
        k_\tau(\mathbf{z},\mathbf{z}^{-})
    \right].
\end{align}
Following~\cite{deng2026generative}, local affinity is measured by the multiscale exponential kernel
\begin{equation}
    k_\tau(\mathbf{z},\mathbf{z}')
    =
    \exp\left(
        -\frac{\left\Vert\mathbf{z}-\mathbf{z}'\right\Vert_2}{\tau}
    \right),
    \label{eq_kernel_exp_base}
\end{equation}
with temperature $\tau$. Combining the attractive and repulsive terms yields the unified field:
\begin{equation}
    \resizebox{\linewidth}{!}{$
    \displaystyle
    \mathbf{V}_{p,q}(\mathbf{z})
    =
    \frac{1}{Z_p(\mathbf{z})Z_q(\mathbf{z})}
    \mathbb{E}_{\mathbf{y}^{+} \sim p, \, \mathbf{y}^{-} \sim q}
    \left[
        k_\tau(\mathbf{z},\mathbf{z}^{+})
        k_\tau(\mathbf{z},\mathbf{z}^{-})
        (\mathbf{z}^{+} - \mathbf{z}^{-})
    \right]
    $}.
    \label{eq_joint_drift_field_base}
\end{equation}

\subsection{Drifting as Score Matching}
\label{ssec:score_difference}

In score-based generative modeling~\cite{song2019generative, weber2023the}, the score function $\nabla_{\mathbf{z}} \log p(\mathbf{z})$ defines a vector field that points in the direction of increasing probability density. Therefore, the score difference between the target data distribution and the model distribution provides an instantaneous corrective velocity that guides generated samples toward higher density regions. When replacing the kernel in~\eqref{eq_kernel_exp_base} with a Gaussian kernel of bandwidth $\sigma$, the drift operator admits a closed-form identity that recovers this scaled score difference~\cite{turan2026generative}:
\begin{equation}
    \mathbf{V}_{p,q}^{(\sigma)}(\mathbf{z}) = \sigma^2 \nabla_{\mathbf{z}} \log p_\sigma(\mathbf{z}) - \sigma^2 \nabla_{\mathbf{z}} \log q_\sigma(\mathbf{z}) = \mathbf{v}_{\sigma}[q](\mathbf{z}).
    \label{eq_score_difference}
\end{equation}
This mathematical identity shows that the drifting field aligns with an instantaneous score-based corrective direction on smoothed densities.

The score matching interpretation conceptually mirrors Distribution Matching Distillation (DMD)~\cite{yin2024one}, which also steers generation by evaluating a score difference. However, DMD parameterizes this with a fake score network and a pretrained diffusion teacher:
\begin{equation}
    \mathbf{V}_{\text{DMD}}(\mathbf{z}) = \mathbb{E}_{t, \epsilon} \left[ \omega(t) \left( s_{\text{teacher}}(\mathbf{z}_t, t) - s_{\text{fake}}(\mathbf{z}_t, t) \right) \right].
\end{equation}
In contrast, generative drifting eliminates the reliance on time-conditioned diffusion teachers by estimating the score difference directly from clean data and model samples in the latent space.

%% file: sections/method.tex
\section{Method}
\label{sec:method}

\begin{figure*}[!htbp]
  \centering
  \includegraphics[width=\textwidth]{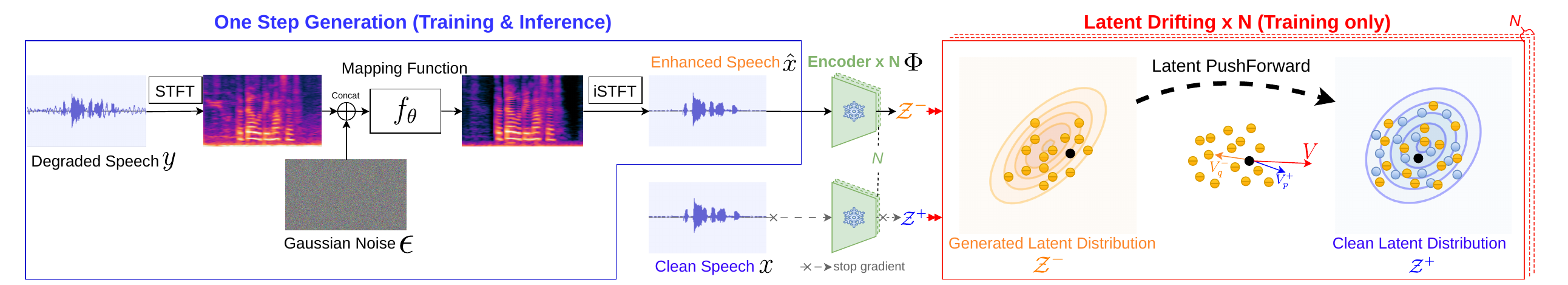}
  \caption{Overview of the DriftSE framework, illustrating its architectural and training flexibility. One-step generation is performed by the mapping network $f_{\theta}$ in a single forward pass by concatenating the degraded speech $\mathbf{y}$ with a fresh Gaussian noise $\boldsymbol{\epsilon}$. During training, the generated speech $\hat{\mathbf{x}}$ and clean speech $\mathbf{x}$ are projected through $N$ encoders (typically $N=2$) into their respective latent spaces to facilitate latent drifting. Within each latent space, an empirical drifting field $\mathbf{V}$ drives the generated distribution $\mathcal{Z}^{-}$ toward the clean distribution $\mathcal{Z}^{+}$. During inference, the encoders and drifting fields are discarded, yielding strictly 1-NFE generation. Crucially, DriftSE natively supports multi-latent drifting and unpaired training.}
  \label{fig:drifting_paradigms}
  \vspace{-1em}
\end{figure*}

We propose DriftSE, a generative framework that formulates speech enhancement as a latent distribution equilibrium problem, extending our previous work~\cite{xu2026driftse}. An overview is illustrated in Fig.~\ref{fig:drifting_paradigms}.

\subsection{One-Step Generation}
\label{ssec:generator}

Let \(\mathbf{y}\) and \(\hat{\mathbf{x}}\) denote the degraded input and enhanced output waveforms, respectively. Following~\cite{will2016complex, richter2023speech, welker2026streamfm}, the mapping network \(f_\theta\) operates in the complex Short-Time Fourier Transform (STFT) domain to jointly model magnitude and phase. Let \(\mathcal{S}(\cdot)\) and \(\mathcal{S}^{-1}(\cdot)\) denote the STFT and inverse STFT operations. The generation process is formulated as
\begin{equation}
  \hat{\mathbf{x}} =
  \mathcal{S}^{-1} \Big(
  f_{\theta}\big([\gamma \boldsymbol{\epsilon}; \mathcal{S}(\mathbf{y})]\big)
  \Big),
  \qquad
  \boldsymbol{\epsilon} \sim \mathcal{N}(\mathbf{0}, \mathbf{I}),
  \label{eq:driftse_generator}
\end{equation}
where \(\boldsymbol{\epsilon}\) is a fresh Gaussian noise sample scaled by a noise level \(\gamma\), and \([\cdot;\cdot]\) denotes channel-wise concatenation. The mapping network \(f_\theta\) therefore receives the scaled noise and degraded complex spectrum as a single input tensor and generates the enhanced complex spectrum in one forward pass.

\subsection{Generative Latent Drifting}
\label{ssec:latent_drifting}

To optimize \(f_\theta\), the drifting field must provide perceptually and acoustically meaningful transport directions. As established in Section~\ref{sec:Background}, evaluating this field directly in raw waveform or STFT domains is ill-defined. We therefore perform distribution alignment entirely within latent spaces, where local neighborhoods capture the underlying speech structure.

\subsubsection{Latent Representations}
\label{sssec:representation_spaces}

To ensure the generated speech is both linguistically intact and physically natural, DriftSE natively aligns distributions across $N$ parallel latent spaces to capture complementary speech properties. We examine semantic latents~\cite{chen2022wavlm, hsu2021hubert, chang2022distilhubert} for phonetic integrity and acoustic latents~\cite{pmlr-v202-chen23ag, kong2020panns} for physical environmental cues. To benefit from both domains, we propose dual-latent drifting to align the two spaces in parallel, and additionally benchmark against joint semantic-acoustic latents~\cite{yang2026wavcube}.

Let \(\Phi_n(\cdot)\) denote the \(n\)-th pretrained encoder, where \(n \in \{1,\dots,N\}\). For an enhanced waveform \(\hat{\mathbf{x}}\) and a clean reference \(\mathbf{x}\), the encoder extracts sequences of \(M_{n}\) frame-level features formulated as
\begin{equation}
    \mathbf{z}^{-}_{n} = \Phi_n(\hat{\mathbf{x}}),
    \qquad
    \mathbf{z}^{+}_{n} = \Phi_n(\mathbf{x}).
\end{equation}
Let \(\mathbf{z}^{-}_{n,b,m}\) and \(\mathbf{z}^{+}_{n,b,m}\) denote the \(m\)-th frame of the \(b\)-th utterance. Aggregating across a mini-batch of size \(B\), we pool all \(B \times M_{n}\) frames to construct the global generated and clean latent sets \(\mathcal{Z}^{-}_{n}\) and \(\mathcal{Z}^{+}_{n}\). These sets serve as the discrete empirical distributions \(q_n\) and \(p_n\) required to compute the latent drifting field.

\subsubsection{Empirical Latent Drifting}
\label{sssec:drifting_objective}

For each latent space, DriftSE computes an empirical estimator of the theoretical drifting field in~\eqref{eq_joint_drift_field_base} by replacing the expectations with sample averages over the pooled mini-batch frames:
\begin{equation}
    \resizebox{0.92\linewidth}{!}{$\displaystyle
    \mathbf{V}_n(\mathbf{z}) = \frac{1}{Z_{p,n}(\mathbf{z}) Z_{q,n}(\mathbf{z})} \sum_{\mathbf{z}^+ \in \mathcal{Z}^+_n} \sum_{\mathbf{z}^- \in \mathcal{Z}^-_n} k_\tau(\mathbf{z}, \mathbf{z}^+) k_\tau(\mathbf{z}, \mathbf{z}^-) (\mathbf{z}^+ - \mathbf{z}^-)
    $}
    \label{eq:joint_drift_field_empirical}
\end{equation}
where \(Z_{p,n}(\mathbf{z})\) and \(Z_{q,n}(\mathbf{z})\) are the empirical normalizers over $\mathcal{Z}^+_n$ and $\mathcal{Z}^-_n$. This kernel-weighted mean-shift force explicitly repels the query frame from the empirical frame-wise density of the current model outputs \(\mathcal{Z}^-_n\) and attracts it toward high-density regions of the clean manifold \(\mathcal{Z}^+_n\).

Instantiating the general latent update in~\eqref{eq_latent_update} at the frame level, we set the query point to each generated frame, $\mathbf{z}_{n,b,m} = \mathbf{z}^{-}_{n,b,m} \in \mathcal{Z}^{-}_n$. The corresponding drifted target is given by
\begin{equation}
    \tilde{\mathbf{z}}_{n,b,m}
    =
    \mathbf{z}_{n,b,m}
    +
    \mathbf{V}_{n}(\mathbf{z}_{n,b,m}),
    \label{eq:latent_target}
\end{equation}
where the empirical velocity field $\mathbf{V}_{n}(\cdot)$ absorbs the step size $h$.
The training objective minimizes the discrepancy between each generated frame and its drifted target, averaged across all pooled frames and balanced by a latent weight \(\lambda_n\):
\begin{equation}
    \mathcal{L}_{\mathrm{drift}}
    =
    \sum_{n=1}^{N}
    \frac{\lambda_n}{B M_n}
    \sum_{b=1}^{B} \sum_{m=1}^{M_n}
    \left\|
    \mathbf{z}_{n,b,m}
    -
    \sg\!\left(
        \tilde{\mathbf{z}}_{n,b,m}
    \right)
    \right\|_2^2,
    \label{eq:drift_loss_generalized}
\end{equation}
where $M_n$ denotes the number of latent frames extracted by encoder $\Phi_n$ from each utterance. For $N=1$, this objective reduces to single-latent drifting~\cite{xu2026driftse}. For $N>1$, it generalizes to multi-latent drifting, aligning distributions across different latents. In this work, we specifically employ dual-latent drifting to simultaneously match distributions within semantic and acoustic spaces. Although in practice the drifting loss is calculated across multiple layers and temperatures, for clarity the training procedure is summarized in Algorithm~\ref{alg:driftse_training} for the single-layer, single-temperature case.

\begin{algorithm}[!htbp]
    \caption{DriftSE Training and Inference}\label{alg:driftse_training}
\begin{algorithmic}[1]
    \State \textbf{Input:} Clean dataset $\mathcal{D}_{x}$, degraded $\mathcal{D}_{y}$, STFT $S(\cdot)$, iSTFT $S^{-1}(\cdot)$, mapping network $f_\theta$, $N$ encoders $\{\Phi_n\}_{n=1}^N$, weights $\{\lambda_n\}_{n=1}^N$, temperature $\tau$, scale prior $p_{\gamma}$
    \State \textbf{Output:} Enhanced speech $\hat{x}_{\mathrm{test}}$
    
    \vspace{4pt}
    \State \textcolor{gray}{\textit{// Training Phase}}
    \Repeat
        \vspace{2pt}
        \State \textcolor{gray}{\textit{// 1. Data Sampling and Generation}}
        \State Sample batches $x \sim \mathcal{D}_{x}$ and $y \sim \mathcal{D}_{y}$         
        \State Sample noise $\epsilon \sim \mathcal{N}(0, I)$ and noise level $\gamma \sim p_{\gamma}$         
        \State Generate batch $\hat{x} \gets S^{-1} \big( f_{\theta}([\gamma \epsilon; S(y)]) \big)$
        \vspace{2pt}
        \State \textcolor{gray}{\textit{// 2. Latent Extraction and Drifting}}
        \State $\mathcal{L}_{\mathrm{drift}} \gets 0$
        \For{$n = 1, \dots, N$}
            \State $\{z^{+}_{n,b,m}\} \gets \Phi_n(x)$ and $\{z^{-}_{n,b,m}\} \gets \Phi_n(\hat{x})$             
            \State Pool $B \times M_n$ frames into $\mathcal{Z}^+, \mathcal{Z}^- \in \mathbb{R}^{B M_n \times d_n}$   
            \State $\mathcal{Z} \gets \mathcal{Z}^-$ \Comment{Current distribution}
            \State $V_n \gets \mathrm{Compute\_V}(\mathcal{Z}, \mathcal{Z}^+, \mathcal{Z}^-, \tau)$ \Comment{See~\protect\eqref{eq:joint_drift_field_empirical}}
            \State Drift target $\tilde{\mathcal{Z}} \gets \mathcal{Z} + V_n$             
            \State $\mathcal{L}_{\mathrm{drift}} \gets \mathcal{L}_{\mathrm{drift}} + \frac{\lambda_n}{B M_n} \big\| \mathcal{Z} - \mathrm{sg}(\tilde{\mathcal{Z}}) \big\|_F^2$
        \EndFor
        
        \vspace{2pt}
        \State \textcolor{gray}{\textit{// 3. Optimization (learning rate $\alpha$)}}
        \State Update $\theta \gets \theta - \alpha \nabla_\theta \mathcal{L}_{\mathrm{drift}}$     
    \Until{convergence}
    
    \vspace{4pt}
    \State \textcolor{gray}{\textit{// Inference Phase}}
    \For{each test batch $y_{\mathrm{test}} \sim \mathcal{D}_{\mathrm{test}}$}
        \State Sample noise $\epsilon \sim \mathcal{N}(0, I)$ and set fixed $\gamma_{\mathrm{inf}}$
        \State Generate $\hat{x}_{\mathrm{test}} \gets S^{-1} \big( f_{\theta}([\gamma_{\mathrm{inf}} \epsilon; S(y_{\mathrm{test}})]) \big)$  
    \EndFor
\end{algorithmic}
\end{algorithm}

\subsection{Unpaired Training}
\label{ssec:unpaired}

DriftSE natively supports fully unpaired training. The empirical drifting field $\mathbf{V}_n$ in~\eqref{eq:joint_drift_field_empirical} is computed over batch-wise latent sets $\mathcal{Z}^-_n$ and $\mathcal{Z}^+_n$ without requiring time-aligned utterance pairs $(\mathbf{y}, \mathbf{x})$. Because evaluating kernel similarities across pooled frames requires no index-level or temporal correspondence, degraded inputs $\mathbf{y} \sim \mathcal{D}_y$ and clean targets $\mathbf{x} \sim \mathcal{D}_x$ can be drawn independently from different speech datasets during training.

%% file: sections/experiments.tex
\section{Experiments}
\label{sec:experiments}

\subsection{Latent Representations}
\label{ssec:latents}

DriftSE can incorporate arbitrary pretrained encoders to construct the parallel latent spaces $\mathcal{Z}_n$. Based on their pretraining objectives, we categorize the evaluated encoders into \emph{semantic}, \emph{acoustic}, and \emph{joint} representations (Table~\ref{tab:encoder_details}).

\paragraph{Semantic latents}
Semantic latents are trained on human speech to capture meaning-bearing units like phonemes and high-level structure. By prioritizing phonetic integrity, they reduce sensitivity to local acoustic variations. HuBERT~\cite{hsu2021hubert} predicts masked cluster IDs obtained from MFCC features~\cite{davis1980comparison}, then iteratively refines these targets using representations from a preceding model. DistilHuBERT~\cite{chang2022distilhubert} inherits this semantic geometry with multilayer knowledge distillation from a HuBERT teacher. WavLM~\cite{chen2022wavlm} extends the objective by predicting the cluster IDs of a clean utterance from inputs mixed with noise or overlapping speech. While we use {semantic} and {linguistic} interchangeably throughout this work, the distinction between structural speech representations and lexical semantic meaning is discussed in~\cite{shi2026speech}.

\paragraph{Acoustic latents}
The acoustic latents employed in this work are trained on general audio targets rather than speech-specific units. Consequently, they emphasize broad environmental patterns without imposing a phonetic organization. PANNs~\cite{kong2020panns} is trained through weakly supervised multilabel audio tagging, which requires its representations to identify discrete sound events. BEATs~\cite{pmlr-v202-chen23ag} predicts masked self distilled token IDs from general audio. This objective encourages the network to capture stable acoustic identities.

\paragraph{Joint latent}
Joint latent representations explicitly combine phonetic structure with fine acoustic detail. WavCube~\cite{yang2026wavcube} first compresses WavLM~\cite{chen2022wavlm} features into a semantic bottleneck, then injects acoustic detail through end-to-end fine-tuning.

\begin{table*}[!htbp]
\centering
\caption{Pretrained encoders, pretraining objectives, and hyperparameters\textsuperscript{\protect\hyperlink{fn:checkpoints}{2}}. Each model extracts frame features from the drifting layers to construct the empirical latent distributions. All extracted layers are weighted equally.}
\label{tab:encoder_details}
\begin{tabular}{l c p{6.5cm} l c c}
\toprule
\textbf{Encoder} & \textbf{Type} & \textbf{Pretraining Objective} & \textbf{Drifting Layers} & \textbf{Dim.} & \textbf{Temperatures ($\tau$)} \\
\midrule
HuBERT~\cite{hsu2021hubert} & Semantic & Masked prediction of intermediate latent cluster IDs & $\{6, 12, 24\}$ & 1024 & $\{0.005, 0.01\}$ \\
DistilHuBERT~\cite{chang2022distilhubert} & Semantic & Knowledge distillation from HuBERT & $\{0, 1, 2\}$ & 768 & $\{0.02, 0.05, 0.1\}$ \\
WavLM~\cite{chen2022wavlm} & Semantic & Masked prediction of clean pseudo-labels from noisy speech & $\{6, 12, 24\}$ & 1024 & $\{0.005, 0.01\}$ \\
\midrule
PANNs~\cite{kong2020panns} & Acoustic & Weakly-supervised multi-label audio tagging & Blocks $\{5, 6\}$ & 1024, 2048 & $\{0.005, 0.01\}$ \\
BEATs~\cite{pmlr-v202-chen23ag} & Acoustic & Masked prediction of self-distilled token IDs from audio & $\{6, 12\}$ & 768 & $\{0.005, 0.01\}$ \\
\midrule
WavCube Pro~\cite{yang2026wavcube} & Joint & WavLM semantic compression + acoustic injection & Final layer & 128 & $\{0.02, 0.05, 0.1\}$ \\
\bottomrule
\end{tabular}
\vspace{-2em}
\end{table*}
\footnotetext[2]{\hypertarget{fn:checkpoints}{}Official checkpoints: HuBERT (\url{https://huggingface.co/facebook/hubert-large-ll60k}), DistilHuBERT (\url{https://huggingface.co/ntu-spml/distilhubert}), WavLM (\url{https://huggingface.co/microsoft/wavlm-large}), PANNs (\url{https://github.com/qiuqiangkong/audioset\_tagging\_cnn}), BEATs (\url{https://github.com/microsoft/unilm/tree/master/beats}), and WavCube Pro (\url{https://huggingface.co/yhaha/WavCube/tree/main/WavCube-pro}).}

\subsection{Datasets}
\label{ssec:datasets}

We evaluate DriftSE across four standard benchmarks spanning two primary tasks: speech denoising and dereverberation. Throughout our experiments, all speech signals are downsampled to 16 kHz.

\paragraph{Speech Denoising} 
The EARS-WHAM dataset mixed clean speech from the EARS~\cite{richter24_interspeech} with non-stationary noise from the WHAM!~\cite{wichern2019wham} at SNRs computed from K-weighted loudness levels uniformly sampled from $[-2.5, 17.5]$ dB.

{VoiceBank-DEMAND (VB-DMD):} The training set comprises clean utterances~\cite{botinhao2016investigating} mixed with eight real-world noises from the DEMAND database~\cite{thiemann2013diverse} and two synthetic noises (babble and speech-shaped) at SNRs of 0, 5, 10, and 15 dB. The test set is generated using unseen noise types at SNRs of 2.5, 7.5, 12.5, and 17.5 dB. We hold out speakers \texttt{p226} and \texttt{p287} from the training data to serve as a validation set as used in~\cite{richter2023speech}.

\paragraph{Speech Dereverberation}
The EARS-Reverb dataset~\cite{richter24_interspeech} is simulated using 2,313 RIRs with diverse characteristics sourced from seven datasets ($RT_{60} \le 2$ s).

{WSJ0-REVERB:} Following~\cite{richter2023speech}, clean WSJ0 utterances~\cite{garofolo1993csrwsj0} are convolved with simulated room impulse responses (RIRs)~\cite{scheibler2018pyroomacoustics} ($T_{60} \in [0.4, 1.0]$ s), yielding an average direct-to-reverberant ratio of $\approx -9$ dB. Anechoic targets are synthesized using matched room geometries with near-total absorption.

\paragraph{Unpaired Clean Corpus} 
To validate the unpaired training capability described in Section~\ref{ssec:unpaired}, we employ the training set from the Deep Noise Suppression Challenge 2020 (DNS2020)~\cite{reddy2020interspeech} as a completely mismatched clean corpus. DNS2020 comprises 500 hours of clean speech from 2,150 speakers, sourced from the LibriVox~\cite{panayotov2015librispeech} audiobook corpus. When conducting unpaired experiments, noisy utterances $\mathbf{y}$ are drawn from the training sets of VB-DMD or WSJ0-REVERB, while the clean targets $\mathbf{x}$ are drawn from DNS2020.

\subsection{Evaluation Metrics}
\label{ssec:metrics}

To comprehensively assess enhancement quality, we employ both intrusive (reference-based) and non-intrusive (reference-free) metrics. Our intrusive evaluation captures downstream linguistic preservation (WER), perceptual quality (PESQ), time-domain signal fidelity (SI-SDR), and speech intelligibility (ESTOI). To evaluate without clean references, we further report a suite of non-intrusive perceptual predictors (DiMOS, WVMOS, NISQA, and SCOREQ). Finally, beyond acoustic quality, we report the computational efficiency through strict system-level profiling.

\paragraph{Intrusive Metrics}
Intrusive metrics require a time-aligned clean reference to evaluate the enhanced speech.
\begin{itemize}
    \item{WER:} Word Error Rate evaluates the preservation of downstream linguistic content. We compute WER using the QuartzNet15x5Base-En model~\cite{kriman2020quartznet} from the NVIDIA NeMo toolkit~\cite{kuchaiev2019nemo}, utilizing transcripts derived from the clean reference audio as ground truth, as proposed in~\cite{welker2026streamfm}.
    \item{PESQ:} Perceptual Evaluation of Speech Quality~\cite{rix2001perceptual} evaluates overall speech quality, scaled within the range [1, 4.5].
    \item{SI-SDR:} Scale-Invariant Signal-to-Distortion Ratio~\cite{le2019sdr} quantifies time-domain waveform reconstruction fidelity while remaining robust to scaling mismatches.
    \item{ESTOI:} Extended Short-Time Objective Intelligibility~\cite{jensen2016algorithm} evaluates human speech intelligibility, scaled within the range [0, 1].
\end{itemize}

\paragraph{Non-Intrusive Metrics}
Non-intrusive metrics directly predict human perceptual quality on a standard $[1, 5]$ Mean Opinion Score (MOS) scale without requiring a ground-truth reference. We employ these predictors to provide a holistic assessment of the enhanced speech, bypassing the strict time-alignment and phase-matching constraints of intrusive metrics like SI-SDR that heavily penalize the natural structural variations produced by generative models.

\begin{itemize}
    \item{DiMOS:} A distilled MOS predictor~\cite{stahl2025distillation} that leverages scalable self-supervised representations for robust speech quality assessment.
    \item{WVMOS:} A MOS predictor~\cite{andreev2023hifipp} built on fine-tuned wav2vec 2.0 representations~\cite{baevski2020wav2vec}, shown to correlate strongly with subjective human quality ratings across diverse acoustic degradations.
    \item{NISQA:} A deep learning model~\cite{mittag2021nisqa} utilizing CNNs and self-attention to predict overall mean opinion scores.
    \item{SCOREQ:} A contrastive-regression-based quality predictor~\cite{ragano2024scoreq} that achieves exceptional domain generalization, ensuring accurate speech evaluation across diverse, unseen acoustic environments.
\end{itemize}

\paragraph{Computational Metrics}
To evaluate the computational efficiency and real-time viability of DriftSE, we report the following system-level profiles:
\begin{itemize}
    \item{Model Parameters (Para):} The total number of trainable network weights, reported in millions (M).
    \item{Number of Function Evaluations (NFE):} The number of neural network forward passes required to generate the enhanced speech.
    \item{Multiply-Accumulate Operations (GMACs):} The number of multiply-add operations required to process a 1-second audio segment, reported in billions (giga-MACs). For iterative models, this is calculated as the baseline GMACs per step multiplied by the NFE.
\end{itemize}

\subsection{Implementation Details}
\label{ssec:implementation}

\paragraph{Backbone architectures}
We instantiate $f_\theta$ by adapting three commonly used backbones in SE to verify flexibility across causal and non-causal settings. For offline evaluation, we employ the NCSN++ U-Net~\cite{richter2023speech} and a non-causal TF-GridNet~\cite{wang2023tf}. To demonstrate compatibility with low-latency requirements, we adapt causal backbones using the SFMUnet~\cite{welker2026streamfm} and a causal TF-GridNet with a 32\,ms algorithmic delay.

Following~\cite{richter2023speech, welker2026streamfm}, we also employ channel-wise early fusion by concatenating the real and imaginary components of the scaled Gaussian noise $\gamma\boldsymbol{\epsilon}$ and the degraded STFT observation $\mathcal{S}(\mathbf{y})$. The hierarchical U-Net architectures (NCSN++ and SFMUnet) further re-inject downsampled inputs at each resolution for progressive conditioning.

\paragraph{Training configuration}
The mapping network operates in the complex STFT domain with a 512-point FFT, a hop size of 256, and a Hann window. During training, the noise level $\gamma$ is sampled from $\log \gamma \sim \mathcal{N}(-3.0, 1.2^2)$, truncated at a maximum of $\gamma_{\text{max}} = 0.15$. At inference, $\gamma_{\mathrm{inf}} = 0.05$. All pretrained encoders remain strictly frozen during training. To capture multiscale representations, features extracted from multiple layers (Table~\ref{tab:encoder_details}) are $L_2$-normalized prior to drift computation and evaluated across multiple drifting temperatures. All extracted layers and latent spaces are weighted uniformly, with multi-latent weights set to $\lambda_n = 1.0$. Although the WavCube family releases both standard and Pro checkpoints, we exclusively utilize the WavCube Pro checkpoint and denote it as WavCube in all subsequent results for brevity. We randomly crop utterances to 2 seconds during training, yielding $M=100$ frames per clip for most encoders. Note that due to its architectural downsampling, PANNs CNN14 yields $M=6$ frames for the same 2-second clip. All models are optimized using AdamW at a learning rate of $5\times10^{-4}$ and trained for 100 epochs with a mini-batch size of $B=8$ on an NVIDIA A6000 GPU.

%% file: sections/results.tex
\section{Results and Discussion}

\label{sec:results}
\begin{table*}[!ht]
\centering
\caption{Mean metrics on EARS-WHAM and EARS-REVERB. \textbf{Causal} indicates a causal architecture, \textbf{Para} denotes parameters in millions, and \textbf{GMACs} denotes total operations (per-step $\times$ NFE). Best within a group in \textbf{bold}.}
\label{tab:main_results}
\resizebox{\textwidth}{!}{
\begin{tabular}{l c c c c c c c c c c c c}
\toprule
& & & \multicolumn{2}{c}{\textbf{Complexity}} & \multicolumn{4}{c}{\textbf{Intrusive Metrics}} & \multicolumn{4}{c}{\textbf{Non-Intrusive Metrics}} \\
\cmidrule(lr){4-5} \cmidrule(lr){6-9} \cmidrule(lr){10-13}
\textbf{Model} & \textbf{Causal} & \textbf{Latent} & \textbf{Para} & \textbf{GMACs} & \textbf{WER}~$\downarrow$ & \textbf{PESQ} & \textbf{SI-SDR} & \textbf{ESTOI} & \textbf{DiMOS} & \textbf{WVMOS} & \textbf{NISQA} & \textbf{SCOREQ} \\
\midrule
\multicolumn{13}{c}{\textit{EARS-WHAM (speech denoising)}} \\
\midrule
Noisy & - & - & - & - & 32.80\% & 1.24 & 5.4 & 0.64 & 2.58 & 1.20 & 1.95 & 2.13 \\
\midrule
SGMSE+~\cite{richter2023speech} & $\times$ & - & 65M & 132.89$\times$60 & 18.65\% & 2.20 & 14.2 & 0.84 & 3.93 & \textbf{2.86} & 3.66 & 3.48 \\
ROSE-CD~\cite{xu2025rosecd} & $\times$ & - & 59.62M & 132.88$\times$1 & \textbf{18.19\%} & \textbf{2.81} & 15.3 & 0.85 & 3.92 & 2.60 & 3.61 & \textbf{3.50} \\
DM-IERM~\cite{Guo2026Close} & $\times$ & - & 67M & 129.7$\times$31 & - & 2.67 & \textbf{17.4} & 0.74 & - & - & - & - \\
FM-Euler4~\cite{welker2026streamfm} & $\times$ & - & 73.7M & 107.18$\times$5 & 18.40\% & 2.41 & 16.1 & \textbf{0.86} & \textbf{4.34} & 2.82 & \textbf{4.50} & - \\
\arrayrulecolor{black!20}\midrule\arrayrulecolor{black}
DriftSE (NCSN++) & $\times$ & PANNs & 59.62M & 132.88$\times$1 & 24.34\% & 2.09 & 0.8 & 0.79 & 3.23 & 2.34 & 3.12 & 2.72 \\
DriftSE (NCSN++) & $\times$ & DistilHuBERT & 59.62M & 132.88$\times$1 & 15.19\% & 2.39 & {12.1} & {0.84} & 4.22 & {3.07} & \textbf{4.07} & 3.82 \\
DriftSE (NCSN++) & $\times$ & DistilHuBERT+PANNs & 59.62M & 132.88$\times$1 & \textbf{14.33\%} & 2.46 & \textbf{13.5} & \textbf{0.85} & 4.11 & \textbf{3.13} & {3.96} & \textbf{3.85} \\
DriftSE (NCSN++) & $\times$ & WavCube & 59.62M & 132.88$\times$1 & 15.35\% & 2.42 & 12.0 & {0.84} & \textbf{4.25} & {3.07} & 4.05 & \textbf{3.85} \\
DriftSE (TF-GridNet) & $\times$ & WavCube & 1.69M & 58.88$\times$1 & 15.48\% & \textbf{2.48} & 11.2 & {0.84} & 4.11 & 3.05 & 3.92 & \textbf{3.85} \\
\midrule
SFM-LRK4~\cite{welker2026streamfm} & \checkmark & - & 52.5M & 144.11$\times$5 & 20.10\% & 2.30 & 14.1 & 0.83 & 3.70 & 2.79 & 4.04 & - \\
\arrayrulecolor{black!20}\midrule\arrayrulecolor{black}
DriftSE (SFMUnet) & \checkmark & PANNs & 24.6M & 144.07$\times$1 & 27.15\% & 1.82 & -41.6 & 0.10 & 3.39 & 2.34 & 3.11 & 2.58 \\
DriftSE (SFMUnet) & \checkmark & DistilHuBERT & 24.6M & 144.07$\times$1 & 19.21\% & 2.05 & \textbf{-31.5} & 0.51 & \textbf{4.08} & \textbf{3.05} & 3.70 & 3.33 \\
DriftSE (SFMUnet) & \checkmark & DistilHuBERT+PANNs & 24.6M & 144.07$\times$1 & 18.67\% & 2.13 & -31.6 & 0.52 & {4.04} & {3.03} & 3.77 & 3.30 \\
DriftSE (SFMUnet) & \checkmark & WavCube & 24.6M & 144.07$\times$1 & 18.94\% & 2.21 & -33.9 & 0.51 & 3.71 & 3.00 & 3.80 & 3.35 \\
DriftSE (TF-GridNet) & \checkmark & WavCube & 1.24M & 41.43$\times$1 & \textbf{18.11\%} & \textbf{2.26} & -33.8 & \textbf{0.53} & 3.80 & 3.01 & \textbf{3.92} & \textbf{3.47} \\
\midrule
\multicolumn{13}{c}{\textit{EARS-REVERB (speech dereverberation)}} \\
\midrule
Reverberant & - & - & - & - & 20.10\% & 1.32 & -16.6 & 0.58 & 3.02 & 2.02 & 2.11 & 3.12 \\
\midrule
SGMSE+~\cite{richter2023speech} & $\times$ & - & 65.59M & 132.89$\times$60 & 17.32\% & 1.95 & -12.6 & 0.76 & 3.48 & 2.16 & \textbf{3.58} & \textbf{2.69} \\
ROSE-CD~\cite{xu2025rosecd} & $\times$ & - & 59.62M & 132.88$\times$1 & 15.78\% & 2.69 & -13.9 & 0.82 & 3.75 & \textbf{2.70} & 3.14 & 2.66 \\
DM-IERM~\cite{Guo2026Close} & $\times$ & - & 67M & 129.7$\times$31 & - & \textbf{3.52} & \textbf{14.2} & \textbf{0.92} & - & - & - & - \\
FM-Euler5~\cite{welker2026streamfm} & $\times$ & - & 38.7M & 107.18$\times$5 & \textbf{11.40\%} & 2.31 & -11.7 & 0.85 & \textbf{3.77} & 2.43 & 3.47 & - \\
\arrayrulecolor{black!20}\midrule\arrayrulecolor{black}
DriftSE (NCSN++) & $\times$ & WavLM+PANNs & 59.62M & 132.88$\times$1 & \textbf{8.91\%} & 2.35 & -9.3 & 0.83 & \textbf{4.13} & \textbf{3.04} & 3.89 & \textbf{3.61} \\
DriftSE (NCSN++) & $\times$ & WavCube & 59.62M & 132.88$\times$1 & 10.28\% & 2.33 & -10.5 & 0.82 & 4.04 & 2.72 & \textbf{4.11} & 3.30 \\
DriftSE (TF-GridNet) & $\times$ & WavLM+PANNs & 1.69M & 58.88$\times$1 & 15.59\% & 2.07 & -8.2 & 0.78 & 3.52 & 2.45 & 3.04 & 3.34 \\
DriftSE (TF-GridNet) & $\times$ & WavCube & 1.69M & 58.88$\times$1 & 9.93\% & \textbf{2.43} & \textbf{-6.8} & \textbf{0.84} & 3.92 & 2.69 & 3.92 & 3.17 \\
\midrule
SFM-LRK5~\cite{welker2026streamfm} & \checkmark & - & 27.9M & 144.11$\times$5 & 15.90\% & 2.05 & -13.5 & 0.79 & 3.68 & 2.48 & 3.67 & - \\
\arrayrulecolor{black!20}\midrule\arrayrulecolor{black}

DriftSE (SFMUnet) & \checkmark & WavLM+PANNs & 24.6M & 144.07$\times$1 & 10.97\% & 2.11 & \textbf{-33.3} & \textbf{0.58} & 3.93 & 2.84 & 3.56 & 3.28 \\
DriftSE (SFMUnet) & \checkmark & WavCube & 24.6M & 144.07$\times$1 & 11.17\% & 2.32 & -34.3 & 0.54 & 4.09 & 2.79 & 4.03 & 3.30 \\
DriftSE (TF-GridNet) & \checkmark & WavLM+PANNs & 1.24M & 41.43$\times$1 & 9.00\% & 2.33 & -33.8 & 0.56 & 3.93 & 2.80 & 3.48 & 3.33 \\
DriftSE (TF-GridNet) & \checkmark & WavCube & 1.24M & 41.43$\times$1 & \textbf{8.85\%} & \textbf{2.71} & -34.5 & 0.56 & \textbf{4.16} & \textbf{2.96} & \textbf{4.07} & \textbf{3.58} \\

\bottomrule
\end{tabular}
}
\vspace{-2em}
\end{table*}

\subsection{Main Results}
Table~\ref{tab:main_results} benchmarks DriftSE against state-of-the-art generative baselines on the EARS~\cite{richter24_interspeech} datasets. We retrain SGMSE+~\cite{richter2023speech} and ROSE-CD~\cite{xu2025rosecd}, while directly reporting results for DM-IERM~\cite{Guo2026Close}, as well as the offline (FM-Euler) and streaming (SFM-LRK) variants from~\cite{welker2026streamfm}.

\paragraph{Speech Denoising (EARS-WHAM)}

\textit{Linguistic Integrity (WER).} Downstream linguistic preservation is fundamentally dictated by the latent space. Drifting with purely acoustic latents (PANNs) causes severe linguistic collapse, yielding WERs of 24.34\% offline and 27.15\% in the causal setting. Single semantic drifting (DistilHuBERT) substantially recovers phonetic content, reducing the WER to 15.19\% offline and 19.21\% causally. Crucially, dual-latent drifting (DistilHuBERT+PANNs) outperforms both single-latent configurations across both U-Net backbones, establishing a new state-of-the-art WER of 14.33\% on offline NCSN++ and lowering the causal WER to 18.67\% on SFMUnet. Furthermore, the joint WavCube latent delivers comparable linguistic preservation, while achieving the lowest causal WER of 18.11\% with the causal TF-GridNet backbone.

\textit{Perceptual Quality.} Dual-latent drifting consistently improves the PESQ over single semantic drifting, increasing from 2.39 to 2.46 on offline NCSN++ and from 2.05 to 2.13 on causal SFMUnet, alongside modest gains in ESTOI. Compared with prior offline approaches, dual-latent drifting outperforms iterative baselines SGMSE+ (2.20) and FM-Euler4 (2.41). While the distilled baseline ROSE-CD attains a higher PESQ (2.81), DriftSE achieves a superior SCOREQ of 3.85 compared to the former's 3.50. For causal backbones, the joint latent WavCube delivers the highest perceptual quality, where both SFMUnet (PESQ of 2.21) and TF-GridNet (PESQ of 2.26) achieve strong, competitive results across distinct architectures.

\paragraph{Speech Dereverberation}
\textit{Linguistic Integrity (WER).} We report both dual-latent drifting (WavLM and PANNs) and joint latent drifting (WavCube) for dereverberation. In the offline setting, the NCSN++ backbone with dual latents achieves an 8.91\% WER, surpassing FM-Euler5 (11.40\%). For the causal backbone, TF-GridNet with WavCube latent attains an 8.85\% WER, outperforming the SFM-LRK5 baseline (15.90\%).

\textit{Perceptual Quality.} For offline backbones, the TF-GridNet model with WavCube latent achieves a PESQ of 2.43, outperforming SGMSE+ (1.95) and FM-Euler5 (2.31), while the NCSN++ model with dual latents achieves a SCOREQ of 3.61 compared to 2.69 for SGMSE+ and 2.66 for ROSE-CD. Although the 31-step DM-IERM yields a PESQ of 3.52 with a cascaded regression stage, DriftSE remains highly competitive as a single-step model. This robust performance extends seamlessly to the causal models, where TF-GridNet with the WavCube latent outperforms SFM-LRK5 in all non-intrusive metrics, achieving a NISQA of 4.07 and a DiMOS of 4.16 compared to the latter's 3.67 and 3.68.

\subsection{Latent Ablation}
\paragraph{Speech Denoising}
We report the latent ablation for speech denoising on VB-DMD in Table~\ref{tab:performance_vb}. By evaluating distinct encoders using the offline NCSN++ backbone, a clear dichotomy emerges. Acoustic encoders lack linguistic constraints and are vulnerable to content hallucination, yielding the two highest WERs of the group (11.44\% for PANNs and 9.33\% for BEATs) alongside the poorest waveform fidelity ($-24.3$\,dB and 10.3\,dB SI-SDR, versus 15.6\,dB for DistilHuBERT). Conversely, semantic encoders like DistilHuBERT and joint representations like WavCube excel at preserving phonetic structures. Dual-latent drifting improves fidelity over its semantic component, raising PESQ from 3.00 to 3.07 and SI-SDR from 15.6\,dB to 16.3\,dB with BEATs, on par with the joint latent WavCube (3.05 PESQ, 14.5\,dB). However, it yields no WER improvement over DistilHuBERT (7.80\%). This is expected on a high-SNR benchmark such as VB-DMD, where the phonetic content is already largely intact and the acoustic latent contributes fidelity rather than linguistic constraint. The benefit of dual drifting is therefore clearer under heavier degradation, as in dereverberation. In the causal setting, our SFMUnet backbone with the DistilHuBERT latent attains a comparable PESQ (2.70 vs.\ 2.72) and a higher DiMOS (4.05 vs.\ 3.88) than SFM-Euler4, using 24.6M parameters at 1 NFE against 52.5M at 5 NFE. Finally, because DistilHuBERT achieves performance highly comparable to HuBERT across intrusive and non-intrusive metrics (e.g., WER 7.80\% vs.\ 7.85\%; PESQ 3.00 vs.\ 2.94) with $13\times$ fewer parameters, we omit HuBERT from subsequent evaluations.

\begin{table*}[htbp!]
\centering
\caption{Mean metrics on VoiceBank-DEMAND for speech denoising. Best within a group in \textbf{bold}.}
\label{tab:performance_vb}
\resizebox{\textwidth}{!}{
\begin{tabular}{l c c c c c c c c c c c c}
    \toprule
    & & & \multicolumn{2}{c}{\textbf{Complexity}} & \multicolumn{4}{c}{\textbf{Intrusive Metrics}} & \multicolumn{4}{c}{\textbf{Non-Intrusive Metrics}} \\
    \cmidrule(lr){4-5} \cmidrule(lr){6-9} \cmidrule(lr){10-13}
    \textbf{Method} & \textbf{Causal} & \textbf{Latent} & \textbf{Para} & \textbf{NFE} & \textbf{WER}~$\downarrow$ & \textbf{PESQ} & \textbf{SI-SDR} & \textbf{ESTOI} & \textbf{DiMOS} & \textbf{WVMOS} & \textbf{NISQA} & \textbf{SCOREQ} \\
    \midrule
    Noisy           & - & -             & -    & -   & 9.36\% & 1.99 & 8.4  & 0.78 & 3.56 & 2.83    & 2.76    & 3.07 \\
    MetricGAN+~\cite{fu2021metricganplus}           & $\times$ & -             & 1.8M    & 1   & 9.56\% & 3.02 & 5.9  & 0.80 & 3.50 & 3.62    & 3.93    & 3.81 \\
    UNIVERSE++~\cite{scheibler2024universeplusplus} & $\times$ & -             & 107.5M    & 8   & 8.72\% & 2.91 & 18.0 & 0.85 & 3.74 & 4.39    & \textbf{4.55}    & \textbf{4.35} \\
    SGMSE+~\cite{richter2023speech}                 & $\times$ & -             & 65M  & 60  & 8.83\% & 2.90 & 16.9 & 0.85 & 3.99 & 4.20    & 4.18    & 4.01 \\
    StoRM~\cite{lemercier2023storm}                 & $\times$ & -             & 55M  & 101 & 9.17\% & 2.93 & 18.8  & \textbf{0.88}  & 4.00    & 4.26    & 4.54    & 4.17 \\
    Thunder~\cite{trachu24_interspeech}             & $\times$ & -             & 65M  & 2   & \textbf{7.35\%} & 2.97 & 19.3  & \textbf{0.88}  & 3.94    & 4.23    & \textbf{4.55}    & 4.24 \\
    ROSE-CD~\cite{xu2025rosecd}                     & $\times$ & -             & 59.62M    & 1   & 7.40\% & 3.49 & 17.8 & 0.87 & 3.76 & \textbf{4.40}    & 4.33    & 4.23 \\        
    SBCTM~\cite{nishigori2025schrodinger}           & $\times$ & -             & 65M    & 1   & 7.80\% & \textbf{3.56} & 12.7 & 0.87 & 3.73 & \textbf{4.40}    & \textbf{4.55}    & 4.34 \\
    MeanFlowSE~\cite{li2026meanflowse}              & $\times$ & -             & 65M    & 1   & 7.46\% & 2.81 & \textbf{19.9} & \textbf{0.88} & 3.88 & 4.30    & 4.47    & 4.25 \\
    FM-Euler4~\cite{welker2026streamfm} & $\times$ & - & 73.7M & 5 & - & 2.86 & 14.1 & 0.86 & \textbf{4.34} & - & - & - \\
    \arrayrulecolor{black!20}\midrule\arrayrulecolor{black}    
    DriftSE (NCSN++) & $\times$ & PANNs & 59.62M   & 1 & 11.44\% & 2.49 & -24.3 & 0.59 & 3.56 & 3.91 & 3.74 & 3.50 \\
    DriftSE (NCSN++) & $\times$ & BEATs & 59.62M   & 1 & 9.33\% & 2.84 & 10.3 & 0.85 & 3.90 & 4.27 & 4.12 & 3.79 \\
    DriftSE (NCSN++) & $\times$ & DistilHuBERT & 59.62M   & 1 & 7.80\% & 3.00 & 15.6 & 0.85 & 3.99 & 4.41 & 4.33 & 4.15 \\
    DriftSE (NCSN++) & $\times$ & HuBERT & 59.62M   & 1 & 7.85\% & 2.94 & 12.5 & 0.84 & 4.01 & 4.40 & \textbf{4.44} & 4.14 \\
    DriftSE (NCSN++) & $\times$ & WavLM & 59.62M   & 1 & 7.59\% & 3.03 & 14.0 & 0.85 & \textbf{4.04} & \textbf{4.44} & 4.30 & \textbf{4.17} \\
    DriftSE (NCSN++) & $\times$ & WavCube & 59.62M   & 1 & 7.86\% & 3.05 & 14.5 & 0.85 & 4.02 & 4.34 & 4.31 & 3.94 \\
    DriftSE (NCSN++) & $\times$ & DistilHuBERT+BEATs & 59.62M   & 1 & 7.88\% & 3.07 & \textbf{16.3} & 0.85 & 3.94 & 4.35 & 4.27 & 4.01 \\
    DriftSE (NCSN++) & $\times$ & DistilHuBERT+PANNs & 59.62M   & 1 & 8.09\% & 3.06 & 15.8 & 0.85 & 3.89 & 4.37 & 4.32 & 3.99 \\
    DriftSE (TF-GridNet) & $\times$ & DistilHuBERT & 1.69M   & 1 & \textbf{6.69\%} & \textbf{3.18} & {16.2} & \textbf{0.86} & 4.00 & 4.40 & 4.27 & 4.04 \\
    DriftSE (TF-GridNet) & $\times$ & WavCube & 1.69M   & 1 & 7.09\% & \textbf{3.18} & 13.7 & \textbf{0.86} & 4.03 & 4.38 & 4.32 & 4.07 \\
    \arrayrulecolor{black!20}\midrule\arrayrulecolor{black}                 
    DriftSE (TF-GridNet)$^{\mathrm{U}}$ & $\times$ & WavCube & 1.69M   & 1 & \textbf{20.55\%} & \textbf{1.82} & \textbf{1.5} & \textbf{0.70} & \textbf{3.47} & \textbf{4.04} & \textbf{4.22} & \textbf{3.77} \\  
    SGMSE+$^{\mathrm{U}}$ & $\times$ & - & 65M   & 60 & 99.44\% & 1.10 & -42.4 & 0.00 & 2.05 & 1.00 & 1.18 & 1.53 \\                     
    \arrayrulecolor{black!20}\midrule\arrayrulecolor{black}     
    DriftSE (TF-GridNet)$^{*}$ & $\times$ & DistilHuBERT & 1.69M   & 1 & 7.59\% & \textbf{3.32} & \textbf{19.7} & \textbf{0.87} & \textbf{4.02} & 4.33 & 4.14 & 3.77 \\      
    RegressSE (TF-GridNet)$^{\dagger}$ & $\times$ & DistilHuBERT & 1.69M   & 1 & \textbf{6.38\%} & 3.31 & 16.9 & \textbf{0.87} & 3.97 & \textbf{4.48} & \textbf{4.25} & \textbf{4.21} \\ 
    \midrule
    SFM-Euler4~\cite{welker2026streamfm}      & \checkmark & - & 52.5M & 5 & - & \textbf{2.72} & \textbf{13.4}  & \textbf{0.85} & \textbf{3.88} & - & - & - \\
    SFM-LRK4~\cite{welker2026streamfm}      & \checkmark & - & 52.5M & 5 & - & \textbf{2.72} & 13.0  & 0.84 & 3.70 & - & - & - \\
    \arrayrulecolor{black!20}\midrule\arrayrulecolor{black}   
    DriftSE (SFMUnet)      & \checkmark & DistilHuBERT & 24.6M & 1 & \textbf{7.48\%} & 2.70 & \textbf{-21.7}  & 0.44 & \textbf{4.05} & \textbf{4.39} & \textbf{4.27} & \textbf{3.97} \\
    DriftSE (TF-GridNet)      & \checkmark & DistilHuBERT & 1.24M & 1 & 7.95\% & \textbf{2.77} & -25.7  & \textbf{0.46} & 4.04 & 4.35 & 4.26 & 3.94 \\
    DriftSE (TF-GridNet)      & \checkmark & WavCube & 1.24M & 1 & 9.61\% & 2.40 & -30.3  & 0.21 & 3.73 & 4.06 & 4.04 & 3.49 \\
    \arrayrulecolor{black!20}\midrule\arrayrulecolor{black}          
    DriftSE (TF-GridNet)$^{*}$      & \checkmark & DistilHuBERT & 1.24M & 1 & 8.24\% & {3.19} & \textbf{19.1}  & 0.85 & \textbf{3.94} & 4.33 & \textbf{4.21} & {3.82} \\  
    RegressSE (TF-GridNet)$^{\dagger}$      & \checkmark & DistilHuBERT & 1.24M & 1 & \textbf{7.11\%} & \textbf{3.27} & 15.8  & \textbf{0.86} & 3.90 & \textbf{4.37} & {4.14} & \textbf{4.11} \\         
    \bottomrule
    \multicolumn{13}{l}{\footnotesize
    $^{\mathrm{U}}$ Unpaired-training variant.
    $^{*}$ Model trained jointly with auxiliary time-domain PESQ and SI-SDR losses~\cite{xu2025rosecd}.
    $^{\dagger}$ Pure frame-wise latent regression variant.} \\
\end{tabular}
}
\vspace{-2em}
\end{table*}

\paragraph{Complex Dereverberation}
Table~\ref{tab:performance_wsj} details latent ablations for complex dereverberation on WSJ0-REVERB. Single-domain latent configurations (purely acoustic or purely semantic) exhibit complementary limitations. Acoustic encoders (PANNs, BEATs) fail to enforce linguistic constraints, resulting in severe content hallucination (17.72\% WER for PANNs, underperforming the unprocessed reverberant speech at 15.53\%). Among single semantic encoders, WavLM outperforms DistilHuBERT (5.31\% vs.\ 5.49\% WER; 2.18 vs.\ 2.03 PESQ), so we choose it as the semantic component for dual-latent drifting, serving as our primary semantic anchor. In contrast to additive denoising, dual-latent drifting here delivers substantial WER reductions over WavLM alone, dropping from 5.31\% to 4.21\% (+BEATs) and 4.01\% (+PANNs), while consistently boosting physical fidelity (PESQ increases from 2.18 to 2.37 and SI-SDR from $-5.2$\,dB to $-3.8$\,dB). This demonstrates that under convolutive smearing, the acoustic latent provides critical physical constraints that assist phonetic recovery. Across backbones, offline TF-GridNet with dual-latent drifting (WavLM+PANNs) achieves 2.14\% WER and 2.53 PESQ with only 1.69M parameters, performing on par with joint WavCube (2.59 PESQ, 3.20\% WER) and outperforming the 60-step SGMSE+ baseline (4.38\% WER). In the causal setting, the dual-latent strategy remains robust. The TF-GridNet backbone attains 3.69\% WER, surpassing all offline diffusion baseline that reports WER.

\begin{figure}[!htbp]
  \centering
  \includegraphics[width=0.70\columnwidth]{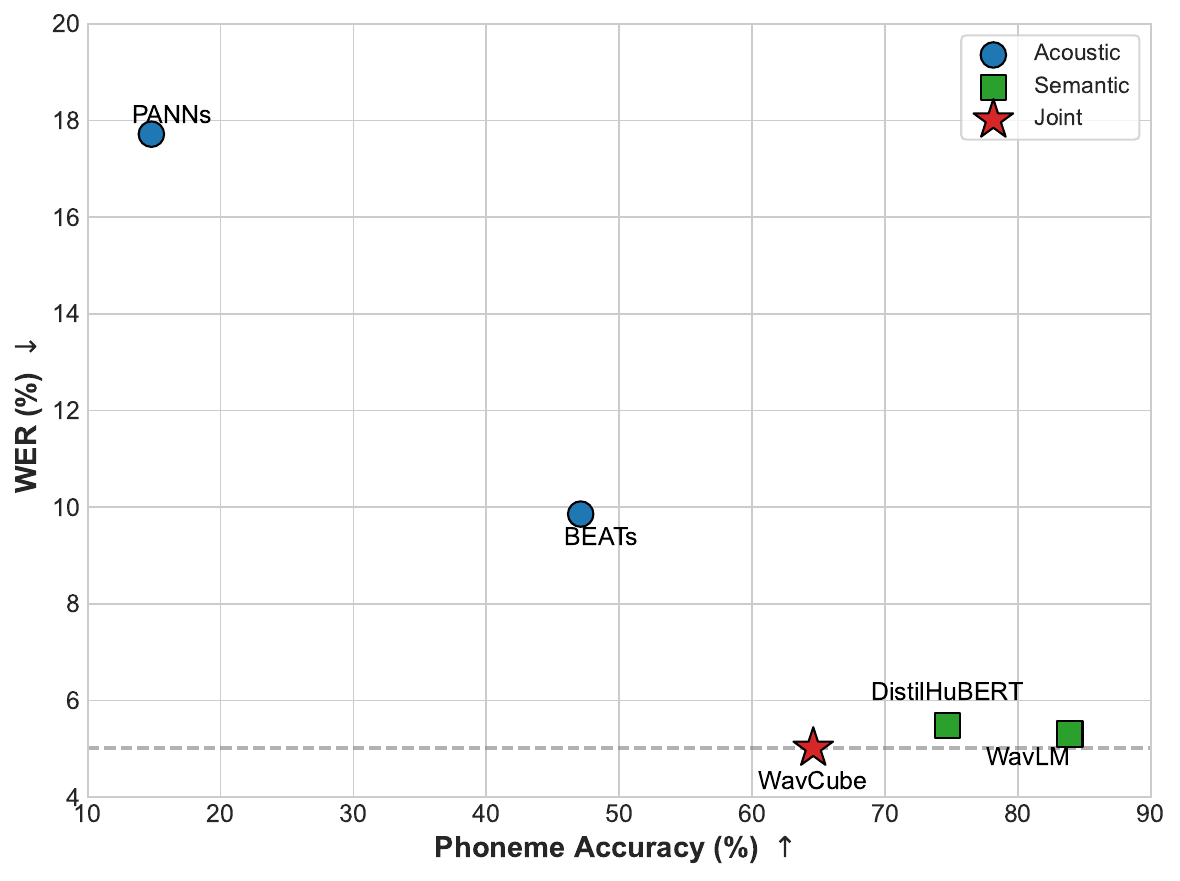}
  \caption{Phoneme accuracy and DriftSE WER.}
  \label{fig:linear_probing}
\end{figure}

\paragraph{Latent Probing}
To measure the intrinsic phonetic separability of different latents, we train a frame-wise linear phoneme classifier~\cite{pasad2021layer} on frozen clean WSJ0-REVERB frames, using 51-class pseudo-phoneme labels generated by a frozen wav2vec~2.0 CTC recognizer~\cite{baevski2020wav2vec}. Fig.~\ref{fig:linear_probing} maps this intrinsic linear phoneme accuracy against the downstream WER of the single-latent DriftSE models in Table~\ref{tab:performance_wsj}.

As illustrated in Fig.~\ref{fig:linear_probing}, intrinsic separability inversely predicts downstream error. Acoustic latents suffer from low phoneme accuracy and high WER, quantitatively confirming their lack of linguistic information and vulnerability to content hallucination. Conversely, semantic latents exhibit high probe accuracy and low WER, demonstrating that their robust phonetic boundaries preserve linguistic integrity. Notably, the joint latent WavCube attains the lowest WER of 5.02\% despite moderate intrinsic separability, suggesting that augmenting semantic structure with acoustic information is beneficial. We validate this by introducing dual-latent drifting in Table~\ref{tab:performance_wsj}, where both the WavLM+BEATs and WavLM+PANNs variants surpass WavLM of 5.31\% and WavCube of 5.02\%, with WavLM+PANNs achieving the best NCSN++ WER of 4.01\%.

\subsection{From Memorization to Generalization}

We compare DriftSE with two regression variants using DistilHuBERT latents and TF-GridNet backbone. DriftSE$^{*}$ is jointly trained with utterance-wise time-domain PESQ and SI-SDR losses~\cite{xu2025rosecd}, while RegressSE$^{\dagger}$ uses pure frame-wise latent regression. Table~\ref{tab:performance_vb} shows that regression improves intrusive metrics. Offline, DriftSE$^{*}$ and RegressSE$^{\dagger}$ increase PESQ from $3.18$ to $3.32$ and $3.31$, respectively. Causally, they increase SI-SDR from $-25.7$\,dB to $19.1$\,dB and $15.8$\,dB. These gains are expected because intrusive metrics reward exact point-to-point memorization, whereas drifting prioritizes latent generalization.

To verify this trade-off~\cite{bamberger2026carr}, we evaluate frame-wise memorization of the VB-DMD training set, alongside generalization on both in-domain (VB-DMD) and cross-domain (EARS-WHAM) test sets. A generated sample $\hat{x}$ is classified as memorized if $M(\hat{x}) = \frac{\|\hat{x} - x^{(1)}\|}{\|\hat{x} - x^{(2)}\|}  \leq\frac{1}{3}$~\cite{yoon2023diffusion}, where $x^{(1)}$ and $x^{(2)}$ are its first and second nearest neighbors in the training set. Generalization is measured with the frame-wise latent Fr\'echet Audio Distance (FAD) against the respective clean test sets. Table~\ref{tab:memorization_generalisation} shows RegressSE$^{\dagger}$ memorizes most and generalizes worst. DriftSE performs best, confirming latent drifting shifts the objective from target memorization to true distribution generalization.
 
\begin{table}[htbp!]
    \centering
    \caption{Frame-wise memorization and generalization evaluated on the last-layer representations of DistilHuBERT.}
    \label{tab:memorization_generalisation}
    \resizebox{0.95\columnwidth}{!}{
    \begin{tabular}{l c cc}
        \toprule
        \multirow{2}{*}{\textbf{Method}} & \multirow{2}{*}{\textbf{Memorization} $\downarrow$} & \multicolumn{2}{c}{\textbf{Generalization (FAD)} $\downarrow$} \\
        \cmidrule(lr){3-4}
        & & \textbf{In-domain} & \textbf{Cross-domain} \\
        \midrule
        RegressSE$^{\dagger}$ & 17.22\% & 1.65 & 8.11 \\
        DriftSE$^{*}$        & 13.33\% & 1.34 & 7.67 \\
        DriftSE              & \textbf{10.68\%} & \textbf{1.16} & \textbf{6.96} \\
        \bottomrule
    \end{tabular}
    }
\end{table}

\begin{table*}[htpb!]
\centering
\caption{Mean metrics on WSJ0-REVERB for speech dereverberation. Best within a group in \textbf{bold}.}
\label{tab:performance_wsj}
\resizebox{\textwidth}{!}{
\begin{tabular}{l c c c c c c c c c c c c}
    \toprule
    & & & \multicolumn{2}{c}{\textbf{Complexity}} & \multicolumn{4}{c}{\textbf{Intrusive Metrics}} & \multicolumn{4}{c}{\textbf{Non-Intrusive Metrics}} \\
    \cmidrule(lr){4-5} \cmidrule(lr){6-9} \cmidrule(lr){10-13}
    \textbf{Method} & \textbf{Causal} & \textbf{Latent} & \textbf{Para} & \textbf{NFE} & \textbf{WER}~$\downarrow$ & \textbf{PESQ} & \textbf{SI-SDR} & \textbf{ESTOI} & \textbf{DiMOS} & \textbf{WVMOS} & \textbf{NISQA} & \textbf{SCOREQ} \\
    \midrule
    Reverberant                     & -      & -            & -     & -   & 15.53\% & 1.31 & -8.6 & 0.45 & 2.76    & 1.49    & 1.78    & 2.69 \\
    SGMSE+~\cite{richter2023speech} & $\times$ & -            & 65M   & 60  & 4.38\% & 2.53 & 3.9  & 0.84 & \textbf{4.38}    & 3.45    & \textbf{4.40}    & 3.59 \\
    StoRM~\cite{lemercier2023storm} & $\times$ & -            & 55M   & 101 & 3.76\% & 2.51 & 6.4  & 0.86 & 4.29    & \textbf{3.69}    & 4.35   & \textbf{3.66} \\    
    ROSE-CD~\cite{xu2025rosecd}     & $\times$ & -            & 59.62M   & 1   & \textbf{3.75\%} & 2.88 & -1.2  & 0.82 &  4.09   & 3.59    & 3.77    & 3.54 \\
    DM-IERM~\cite{Guo2026Close}                         & $\times$ & -            & 67M   & 31  & - & \textbf{3.09} & \textbf{9.9}  & \textbf{0.90} & -    & -    & -    & - \\
    \arrayrulecolor{black!20}\midrule\arrayrulecolor{black}    
    DriftSE (NCSN++)                & $\times$ & PANNs  & 59.62M   & 1   & 17.72\% & 1.61 & -12.3 & 0.68 & 2.71 & 2.51 & 2.90 & 2.20 \\
    DriftSE (NCSN++)                & $\times$ & BEATs  & 59.62M   & 1   & 9.86\% & 1.84 & -31.7 & 0.73 & 3.94 & 2.72 & 3.69 & 2.88 \\
    DriftSE (NCSN++)                & $\times$ & DistilHuBERT   & 59.62M   & 1   & 5.49\% & 2.03 & -5.3 & 0.77 & {4.38} & {3.63} & {3.59} & {3.54} \\
    DriftSE (NCSN++)                & $\times$ & WavLM  & 59.62M   & 1   & 5.31\% & 2.18 & -5.2 & 0.77 & 4.34 & 3.63 & 3.39 & \textbf{3.69} \\    
    DriftSE (NCSN++)                & $\times$ & WavLM+BEATs  & 59.62M   & 1   & 4.21\% & 2.37 & -4.1 & 0.82 & \textbf{4.39} & {3.64} & 3.96 & {3.54} \\    
    DriftSE (NCSN++)                & $\times$ & WavLM+PANNs  & 59.62M   & 1   & 4.01\% & 2.36 & -3.8 & 0.80 & \textbf{4.39} & 3.61 & 3.93 & 3.49 \\
    DriftSE (NCSN++)                & $\times$ & WavCube   & 59.62M   & 1   & 5.02\% & 2.30 & -4.3 & 0.80 & 4.34 & 3.32 & \textbf{4.20} & 3.28 \\    
    DriftSE (TF-GridNet)                & $\times$ & WavLM+BEATs  & 1.69M   & 1   & 2.90\% & 2.43 & -3.4 & \textbf{0.83} & 4.35 & \textbf{3.68} & 3.97 & 3.31 \\    
    DriftSE (TF-GridNet)                & $\times$ & WavLM+PANNs  & 1.69M   & 1   & \textbf{2.14\%} & 2.53 & -3.4 & \textbf{0.83} & 4.25 & 3.58 & 3.80 & 3.43 \\
    DriftSE (TF-GridNet)                & $\times$ & WavCube  & 1.69M   & 1   & 3.20\% & \textbf{2.59} & \textbf{-1.2} & \textbf{0.83} & 4.25 & 3.41 & 4.18 & 3.34 \\    
    \arrayrulecolor{black!20}\midrule\arrayrulecolor{black}        
    DriftSE (TF-GridNet)$^{\mathrm{U}}$                & $\times$ & WavCube  & 1.69M   & 1   & \textbf{15.27\%} & \textbf{1.62} & \textbf{-7.7} & \textbf{0.67} & \textbf{3.93} & \textbf{3.42} & \textbf{4.35} & \textbf{3.05} \\    
    SGMSE+$^{\mathrm{U}}$ & $\times$ & - & 65M   & 60 & 99.81\% & 1.20 & -41.6 & 0.00 & 1.75 & 1.00 & 1.10 & 1.69 \\        
    \midrule              
    DriftSE (SFMUnet)               & \checkmark & WavLM+PANNs  & 24.6M & 1   & 5.34\% & 2.04 & \textbf{-33.0} & \textbf{0.53} & 3.91 & 3.40 & 3.35 & 2.95 \\
    DriftSE (TF-GridNet)               & \checkmark & WavLM+PANNs  & 1.24M & 1   & \textbf{3.69\%} & \textbf{2.18} & -33.3 & 0.52 & \textbf{3.94} & \textbf{3.67} & \textbf{3.71} & \textbf{3.00} \\
    \bottomrule
    \multicolumn{13}{l}{\footnotesize $^{\mathrm{U}}$ Unpaired-training variant.} \\    
\end{tabular}
}
\vspace{-1em}
\end{table*}

\begin{figure*}[!htbp]
  \centering
  \includegraphics[width=1.00\textwidth]{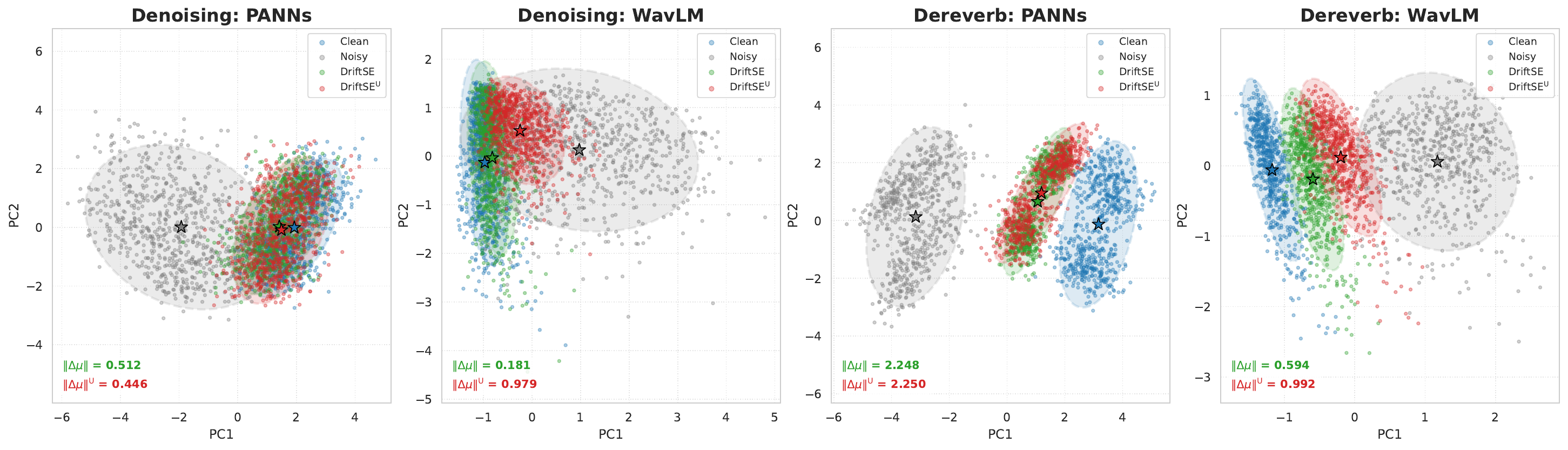}
  \caption{Utterance-level PCA visualizations of DriftSE and DriftSE$^{\mathrm{U}}$ for denoising (left) and dereverberation (right). Both models align closely in the acoustic PANNs space, but the unpaired variant diverges significantly in the semantic WavLM space. $\star$ denotes cluster centroids, and $|\Delta\mu|$ quantifies the distance between generated and clean centroids.}
  \label{fig:unpaired_pca}
\end{figure*}

\subsection{Unpaired Training}

To validate the generative nature of drifting, we conduct fully unpaired training mapping degraded speech (VB-DMD for denoising, WSJ0-REVERB for dereverberation) to an independent clean speech distribution (DNS2020). We report two models in Tables~\ref{tab:performance_vb} and ~\ref{tab:performance_wsj}:

\begin{itemize}
    \item \textbf{DriftSE$^{\mathrm{U}}$:} Trained by independently sampling degraded audio $y \sim \mathcal{D}_{\mathrm{degraded}}$ to generate $\mathcal{Z}^{-} = \Phi(\hat{x})$ and clean audio $x \sim \mathcal{D}_{\mathrm{DNS2020}}$ to extract $\mathcal{Z}^{+} = \Phi(x)$, with a frame batch size of 10{,}240 for stable estimation.
    \item \textbf{SGMSE+$^{\mathrm{U}}$:} Trained using randomly paired degraded $y$ and clean utterances $x_0$ sampled from different datasets.
\end{itemize}

\paragraph{Paired Diffusion versus Drifting}
SGMSE+ defines its forward process using a paired perturbation kernel that interpolates between $\mathbf{x}_0$ and $\mathbf{y}$~\cite[eq.~(8)]{richter2023speech}. For SGMSE+$^{\mathrm{U}}$, randomly pairing unrelated $x_0$ and $y$ renders these diffusion trajectories inconsistent, leading to model collapse. In contrast, DriftSE$^{\mathrm{U}}$ enables unpaired learning because the empirical drifting field aligns aggregate frame distributions rather than enforcing point-to-point temporal trajectories. Even when paired clean utterances are unavailable in the current mini-batch, pooling frame-level latents allows each generated frame to compute kernel affinities across $\mathcal{Z}^{+}$, providing valid mean-shift gradients toward local centers of mass on the clean manifold.

\paragraph{Acoustic Generalization}
DriftSE$^{\mathrm{U}}$ attains high non-intrusive perceptual scores, achieving a NISQA of 4.22 and a SCOREQ of 3.77 for denoising, and a NISQA of 4.35 and a SCOREQ of 3.05 for dereverberation. As illustrated in Fig.~\ref{fig:unpaired_pca}, vanilla DriftSE and DriftSE$^{\mathrm{U}}$ align closely with clean speech in the acoustic PANNs space for denoising ($|\Delta\mu| \le 0.512$). For dereverberation, although both models exhibit a larger offset from the clean centroid ($|\Delta\mu| \approx 2.25$), DriftSE$^{\mathrm{U}}$ tracks DriftSE almost identically ($2.250$ vs.\ $2.248$). This demonstrates that unpaired clean latent drifting still provides meaningful gradient to capture universal acoustic characteristics.

\paragraph{Semantic Limitation}
Unpaired training severely degrades intrusive metrics like WER, increasing from 7.09\% to 20.55\% for denoising and from 3.20\% to 15.27\% for dereverberation. As illustrated in Fig.~\ref{fig:unpaired_pca}, DriftSE$^{\mathrm{U}}$ diverges significantly from the clean distribution in the semantic WavLM space. Because the independent clean mini-batch $\mathcal{Z}^{+}$ lacks corresponding phonetic ground truth, the model drifts frames toward the nearest but incorrect neighbors, inducing severe semantic hallucinations. Thus, while unpaired drifting successfully recovers global acoustic structures, utterance-level pairing remains essential for semantic fidelity.

%% file: sections/conclusion.tex
\section{Conclusion}
\label{sec:conclusion}

This paper presented DriftSE, which formulates speech enhancement as a latent distribution equilibrium problem, achieving strict one-step generation by discarding the drifting field at inference. DriftSE supports both single-latent and dual-latent drifting, and validation confirms that dual-latent drifting across semantic and acoustic representations preserves both speech intelligibility and acoustic quality. Furthermore, the frame-wise drifting objective supports unpaired training without paired noisy-clean data. Experimental evaluations on both additive noise and convolutive reverberation show that DriftSE generalizes across offline and causal backbones, achieving state-of-the-art word error rates alongside competitive perceptual quality.

%% file: config/bios.tex
\vspace{-3em}
\begin{IEEEbiography}[{\includegraphics[width=1in,height=1.25in,clip,keepaspectratio]{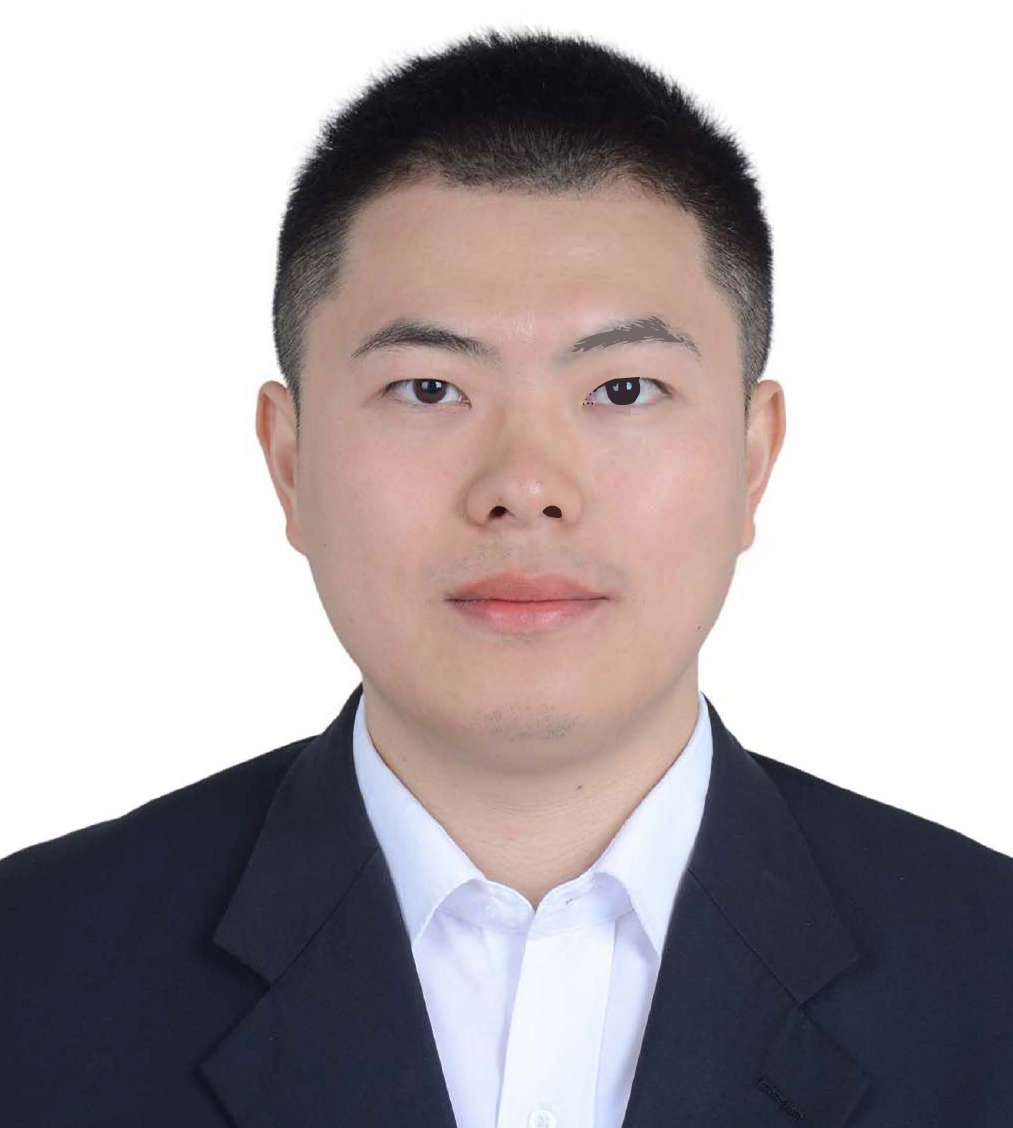}}]{Liang Xu} 
(Student Member, IEEE) received the B.S. degree in Biomedical Engineering from Xidian University, China, in 2016, and the M.S. degree in Information and Communication Engineering from Huazhong University of Science and Technology (HUST), China, in 2019. He is currently pursuing the Ph.D. degree with the School of Engineering and Computer Science, Victoria University of Wellington, New Zealand. His research interests include signal processing, machine learning, and generative models for speech enhancement.
\end{IEEEbiography}%
\vspace{-3em}
\begin{IEEEbiography}[{\includegraphics[width=1in,height=1.25in,clip,keepaspectratio]{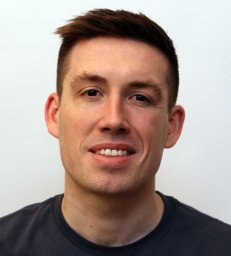}}]{Diego Caviedes-Nozal}
is a Senior Research Scientist at GN Advanced Science, Denmark, where he works on machine learning and generative models for speech and audio. He received the B.Sc. and M.Sc. degrees in telecommunications engineering from the University of Valladolid, Spain, in 2014, and the M.Sc. and Ph.D. degrees in engineering acoustics from the Technical University of Denmark (DTU), in 2016 and 2020, and was a Postdoctoral Researcher with DTU's Acoustic Technology Group in 2021.
\end{IEEEbiography}
\vspace{-3em}
\begin{IEEEbiography}[{\includegraphics[width=1in,height=1.25in,clip,keepaspectratio]{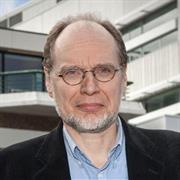}}]{W. Bastiaan Kleijn}
(Life Fellow, IEEE) received the Ph.D. degree in soil science and the M.Sc. degree in physics from the University of California, Riverside, CA, USA, the M.S.E.E. degree from Stanford University, Stanford, CA, USA, and the Ph.D. degree in electrical engineering from TU Delft, Delft, Netherlands. He was a Member of Technical Staff in the Research Division, AT\&T Bell Laboratories. Since 2010, he has been a Professor with Victoria University of Wellington, Wellington, New Zealand, and has also been a Research Scientist with Google since 2011. From 2011 to 2021, he was a Professor with TU Delft, and he was also a Professor at KTH Stockholm from 1996 until 2014. He is a Fellow of the Royal Society of New Zealand and a Fellow of Engineering New Zealand.
\end{IEEEbiography}%
\vspace{-3em}
\begin{IEEEbiography}[{\includegraphics[width=1in,height=1.25in,clip,keepaspectratio]{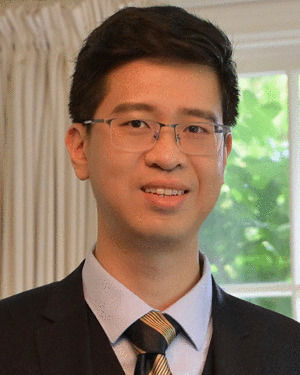}}]{Longfei Felix Yan}
(Member, IEEE) received the B.Sc. (with Hons.) degree from Victoria University of Wellington (VUW), Wellington, New Zealand, in 2017, and the dual Ph.D. degree from VUW and Australian National University, Canberra, ACT, Australia, in 2024. He is currently a Lecturer with Lincoln University, New Zealand. His research interests include statistical signal processing, machine learning, and combinatorial optimization.
\end{IEEEbiography}%
\vspace{-3em}
\begin{IEEEbiography}[{\includegraphics[width=1in,height=1.25in,clip,keepaspectratio]{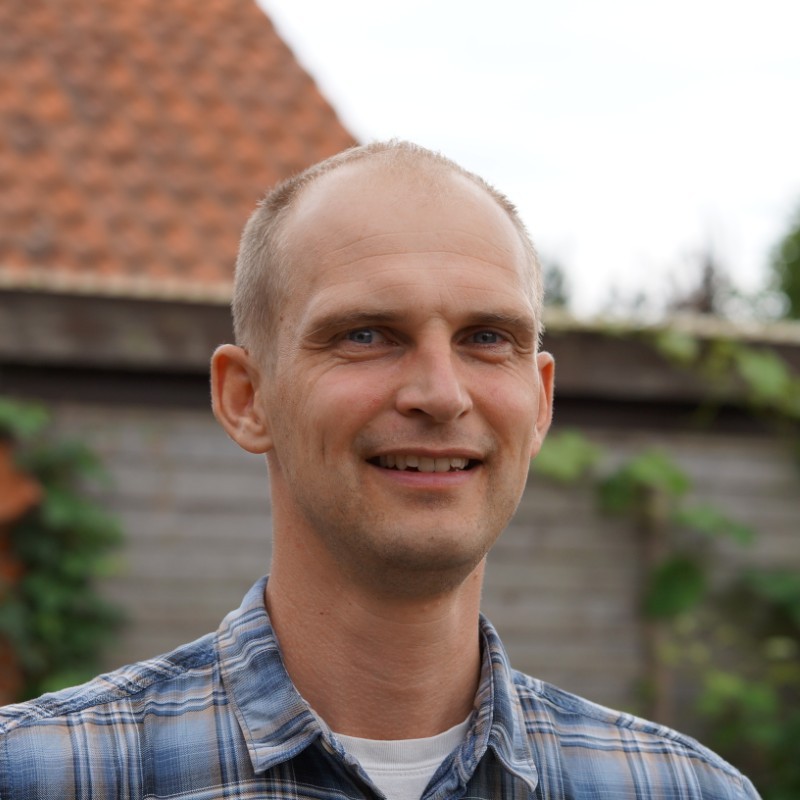}}]{Rasmus Kongsgaard Olsson}
received the Ph.D. degree from the Technical University of Denmark (DTU). He is currently a Principal Research Scientist with GN Advanced Science, Denmark. His research interests include the application of machine learning to acoustic signal-processing tasks.
\end{IEEEbiography}